\documentclass[fleqn,usenatbib]{mnras}

\usepackage[T1]{fontenc}
\usepackage{xcolor}
\usepackage{afterpage}

\DeclareRobustCommand{\VAN}[3]{#2}
\let\VANthebibliography\thebibliography
\def\thebibliography{\DeclareRobustCommand{\VAN}[3]{##3}\VANthebibliography}

\usepackage{graphicx}	
\usepackage{amsmath}	
\usepackage{amssymb}	
\usepackage{float}
\usepackage{placeins}

\title[The magnetic field of V352\,Peg]{The magnetic field of the chemically peculiar star V352\,Peg}

\author[L. Fréour et al.]{
L. Fréour,$^{1,2,3}$
C. Neiner,$^{1}$\thanks{E-mail: coralie.neiner@obspm.fr}
J. D. Landstreet,$^{4,5}$
C. P. Folsom,$^{6}$
G. A. Wade$^{7}$
\\
$^{1}$LESIA, Paris Observatory, PSL University, CNRS, Sorbonne University, Université de Paris, 5 place Jules Janssen, 92195 Meudon, France\\
$^{2}$Central School of Nantes, 1 rue de la Noë, 44300 Nantes, France\\
$^{3}$Technical University of Denmark, Anker Engelunds Vej 1 Bygning 101A, 2800 Kgs. Lyngby, Denmark \\
$^{4}$University of Western Ontario, Department of Physics \& Astronomy, London, Ontario, Canada N6A 3K7\\
$^{5}$Armagh Observatory and Planetarium, College Hill, Armagh BT61 9DG, Northern Ireland, UK\\
$^{6}$University of Tartu, Faculty of Science and Technology, Tartu Observatory, Estonia\\
$^{7}$Dept. of Physics \& Space Science, Royal Military College of Canada, PO Box 17000 Station Forces, Kingston, ON, Canada K7K 0C6
}

\date{Accepted XXX. Received YYY; in original form ZZZ}

\pubyear{2021}

\begin{document}
\label{firstpage}
\pagerange{\pageref{firstpage}--\pageref{lastpage}}
\maketitle

\begin{abstract}
We present a spectropolarimetric analysis of the hot star V352\,Peg.
We have acquired 18 spectropolarimetric observations of the star with ESPaDOnS at the CFHT between 2018 and 2019 and completed our dataset with one archival ESPaDOnS measurement obtained in 2011. Our analysis of the spectra shows that the star is on the main sequence and chemically peculiar, i.e. it is a Bp star, with overabundances of iron peak elements (Ti, Cr and Fe) and underabundance of He and O. Through a Least-Square Deconvolution of each spectrum, we extracted the mean Zeeman signature and mean line profile of thousands of spectral lines and detected a magnetic field in V352\,Peg. By modelling the Stokes I and V profiles and using the Oblique Rotator Model, we determined the geometrical configuration of V352\,Peg. We also performed Zeeman-Doppler Imaging (ZDI) to provide a more detailed characterization of the magnetic field of V352\,Peg and its surface chemical distributions. We find a magnetic field that is mainly dipolar, dominantly poloidal, and largely non-axisymmetric with a dipole field strength of $\sim$9 kG and a magnetic axis almost perpendicular to the rotation axis.
The strong variability of Stokes I profiles also suggests the presence of chemical spots at the stellar surface.
\end{abstract}




\section{Introduction}
Magnetic fields are known to exist in about 10\% of hot stars on the main sequence. 
Recent projects and international collaborations, among which the MiMeS \citep[Magnetism in Massive Stars]{wade2016} and LIFE \citep[the Large Impact of magnetic Fields on the Evolution of hot stars]{martin2018} programmes, have enabled researchers to progress in the understanding of magnetism.

Contrary to cool stars, hot stars exhibit a radiative envelope where a solar-like dynamo cannot arise. The magnetic field present in hot stars is believed to have a fossil origin, most likely a remnant of the magnetic field of the molecular cloud where the star formed \citep[][and references therein]{2001Moss,Neiner2015} or of the field generated during the convective phase of the pre-main sequence \citep[][and references therein]{Alecianetal19},  or possibly resulting from a merger \citep{schneider2016}.
A high percentage of magnetic hot stars also exhibit inhomogenous surface distribution of chemical elements and are classified as chemically peculiar (CP). Through atomic diffusion, chemical elements, channeled by the magnetic field lines, gather in spots distributed around the stellar surface \citep{Michaud1981}. Reconstructing the magnetic field and surface distribution of chemical elements of such stars can help understanding the interaction between chemical elements and magnetic field.

V352\,Peg (= HD\,221218, HIP\,115991) is an $\alpha^2$\,CVn type of star, first detected as a photometrically variable star by the Hipparcos mission \citep{1997ESA}. It has been classified as a B8/9III star in the Hipparcos and Skiff catalogs \citep{1999Skiff}, thus as an evolved hot star. However, more recent observations obtained for the Revised TESS Input Catalog \citep[TIC,][]{Stassun_2019} classified V352\,Peg as a dwarf star.

Here we present high-resolution spectropolarimetric observations of V352\,Peg (Sect.~\ref{obs}). We determine the physical parameters of the star (Sect.~\ref{subsec:stellar_param}) and its chemical composition (Sect.~\ref{subsec:chemical_compo}). We then perform a spectropolarimetric analysis (Sect.~\ref{sec4}) to check for the presence of a magnetic field and determine its strength and geometrical configuration using a Zeeman-Doppler Imaging code developed by \cite{Folsom2018}. We also produce chemical maps of the most abundant elements. Finally, our results and hypotheses are discussed in Sect.~\ref{sec:Discussion}.

\begin{table*}
	\centering
	\caption{Summary of observations, taken with the ESPaDOnS spectropolarimeter at the CFHT. The columns indicate the ESPaDOnS odometer number, the observation date, the Heliocentric Julian Date at the middle of the observation (mid-HJD), the observation universal time, the total exposure time in seconds ($T_{\rm exp}$), the signal to noise ratio, the longitudinal magnetic field ($B_{l}$) with its error bars, and the rotation phase of the observations with respect to the reference ephemeris HJD$_0$ = 2458348.4725 and $P_{\rm rot} = 2.63654$ days.}
	\label{tab:observations}
	\begin{tabular}{lccccccr} 
		\hline
		Espadons odometer & Date & mid-HJD & UT & Exposure & S/N & $B_l \pm \sigma B_l$ & Phase \\
		number &  &   &  hh:mm:ss  & (s)   & per spec pxl & (G) &    \\
		\hline
		1331348 & August 20, 2011  & 2455793.880 & 08:46:04 & 4x410 & 537 & $-2486.2 \pm 12.5$ & 0.074 \\
		2297769 & August 18, 2018 & 2458348.876 & 08:48:41 & 4x195 & 158 & $-2109.9 \pm 381.6$ & 0.153\\
		2297797 & August 18, 2018 & 2458349.079 & 13:41:02 & 4x195 & 414 & $-882.4 \pm 22.5$ & 0.230 \\
		2306359&September 29, 2018 & 2458390.762 & 06:02:41 & 4x195 & 395 & $-2477.7 \pm 15.2$ & 0.040  \\
		2306807 &October 02, 2018 & 2458393.712 & 04:49:55 & 4x195 & 422 & $-1864.1 \pm 18.7$ & 0.159 \\
		2306827& October 02, 2018 & 2458393.931 & 10:06:29 & 4x195 & 416 & $-654.2 \pm 23.2$ & 0.242 \\
		2328690 &October 18, 2018 & 2458409.888 & 09:03:27 & 4x195 & 448 & $972.8 \pm 22.5$ & 0.294\\
		2329187& October 22, 2018 & 2458413.810 & 07:12:20 & 4x195 & 429 & $-1461.5 \pm 21.2$ & 0.782 \\
		2329459&October 23, 2018 & 2458414.814 & 07:18:42 & 4x195 & 458 & $-1838.1 \pm 18.4$ & 0.163\\
		2335338 &November 19, 2018 & 2458441.799 & 06:59:01 & 4x195 & 185 & $2219.0 \pm 28.4$ & 0.398\\
		2335342&November 19, 2018 & 2458441.810 & 07:14:32 & 4x195 & 143 & $2241.2 \pm 55.8$ & 0.402\\
		2363446&December 22, 2018 & 2458474.759 & 06:05:41 & 4x195 & 437 & $-2493.6 \pm 13.2$ & 0.899\\
		2363609&December 23, 2018 & 2458475.687 & 04:21:39 & 4x195 & 357 & $-505.0 \pm 22.7$ & 0.251\\
		2363789&December 24, 2018 & 2458476.698 & 04:37:48 & 4x195 & 422 & $1781.2 \pm 17.0$ & 0.634\\
		2363999&December 25, 2018 & 2458477.716 & 05:03:42 & 4x195 & 385 & $-2462.6 \pm 13.1$ & 0.020\\
		2424591&June 08, 2019 & 2458643.120 & 14:47:43 & 4x180 & 416 & $874.2 \pm 20.6$ & 0.756 \\
		2425097&June 11, 2019 & 2458646.015 & 12:15:58 & 4x180 & 347 & $-2377.4 \pm 16.3 $ & 0.854 \\
		2425307&June 12, 2019 & 2458647.074 & 13:42:09 & 4x180 & 396 & $12.8 \pm 21.3$ & 0.256 \\
		\hline
	\end{tabular}
\end{table*}

\section{Observations}
\label{obs}

Seventeen observations of V352\,Peg were collected between the 17th of August and 11th of June 2019, within the scope of the LIFE project.
An archival observation obtained on the 19th of August 2011, as part of the MiMeS international program is also available.
All the observations were taken with the ESPaDOnS spectropolarimeter at the Canada–France–Hawaii Telescope (CFHT) and cover a large wavelength range, from 370 to 1040 nm, with a spectral resolution of 65\,000.

Each observation consists of 4 sub-exposures, each of them taken in a specific polarimeter configuration. The data reduction pipeline of ESPaDOnS is Libre-Esprit \citep{Donati1997} included in the Upena package. We normalized the spectra through a semi-automatic Python code \citep{Martin2017}.
The Signal to Noise ratio (S/N) of the  spectra ranges from 143 to 537 with an average value of 376, measured at 503 nm. Relevant information regarding the observations is gathered in Table ~\ref{tab:observations}.
Note that, when checking the Night Reports from the CFHT, we noticed large differences in the SNR between sub-exposures for the observation \#2 (2297769). This is not the case for the other observations and could be the evidence of some transient (unknown) issue. There are no obvious systematic errors in the observation so we decided to keep it in the longitudinal magnetic field analysis. However, the Zeeman-Doppler Imaging is more sensitive to small details in line profiles and we rejected observation \#2 from that analysis, since it does not have a significant impact on the rotation phase coverage.

The circular polarization (Stokes V) and null spectra N are computed by respectively constructive and destructive combination of the 4 polarimetric sub-exposures. The total intensity (Stokes I) is also extracted.

\section{Stellar parameters}

\subsection{Fundamental parameters} \label{subsec:stellar_param}

First estimates of the physical parameters (effective temperature $T_{\rm eff}$, luminosity $L$, gravity $\log g$, radius $R$, and mass $M$) of V352\,Peg were obtained from the TIC \citep{Stassun_2019}, based on Gaia DR2 colours and distance. These data are presented in Table\,\ref{tab:stellar_param}. 
Only the error on the effective temperature is available in the TIC. 
We computed the error on the radius by propagating the errors on Gaia distance D, Gaia magnitude G, and $T_{\rm eff}$ in Eq.~(4) from \citet{Stassun_2019}. 
The mass is computed using a unified spline relation between the effective temperature and the mass. The corresponding nodal points are presented in Table 5 from \citet{Stassun_2019}. The uncertainty in the mass is then calculated by propagating the error on $T_{\rm eff}$ in the spline function.
Finally, we estimated the errors on the luminosity and the surface gravity by propagating the errors from the above mentioned parameters in the formula used by \citet{Stassun_2019}.

As a check on the fundamental parameters provided in the TIC, we retrieved the Geneva six-colour photometry for this star from the Lausanne General Catalogue of Photometric Data \citep{Meretal97}, which is maintained online at the Masaryk University in Brno, Czech Republic\footnote{\url{https://gcpd.physics.muni.cz/}}. (No Str\"{o}mgren photometry of this star appears to be available.) These data were used to obtain values of $T_{\rm eff} = 11376 \pm 57$\,K and $\log g = 4.22 \pm 0.08$, employing the program {\sc calib.f} \citep{Kunetal97}, as described in more detail by \citet{SilLan14}. The underlying calibrations for fundamental parameters derived from Geneva photometry with {\sc calib.f} are described by \citet{Kunetal97}. As we will show in Sect.~\ref{subsec:chemical_compo}, V352\,Peg is a chemically peculiar star. {\sc calib.f} does not include options for chemically peculiar stars, but it does allow the user to choose an overall atmospheric abundance of solar, or of solar enhanced or diminished by 1\,dex. We have used the $+1$\,dex enhanced calibrations, but the result changes the computed value of $T_{\rm eff}$ from that of a solar abundance star by only about 100\,K, and $\log g$ changes by only about 0.02\,dex. The uncertainties reported by {\sc calib.f} are fitting uncertainties, which we have replaced with larger uncertainties that we consider to be more realistic for ApBp stars. The agreement of the values obtained from Geneva photometry with that found in the TIC catalog is satisfactory, but we suggest that the TIC uncertainties may be somewhat underestimated for this chemically peculiar star. From the gravity values, it is clear that (in spite of luminosity classification {\sc iii}) V352\,Peg is a main sequence star. 

The rotation period determined from the more than 140 photometric observations  obtained by Hipparcos \citep{Hipparcos1997} is reported in various on-line versions of the Hipparcos variability catalog\footnote{\url{https://vizier.u-strasbg.fr/viz-bin/VizieR-5?-ref=VIZ60e2c396c853e&-out.add=.&-source=I/239/hip_va_1&recno=2660}} to be either $P_{\rm rot} = 2.64$ or $2.63641 \pm 0.00006$\,d. This photometric period will be considerably improved below; Table\,\ref{tab:stellar_param} reports our final value. 

Finally, we estimate the projected rotation velocity $v\sin i$ between 35 km/s and 40 km/s as described below.

\begin{table} 
	\centering
	\caption{Stellar parameters of V352\,Peg. Column 2 indicates the value taken from the reference in Col. 3, while Col. 4 shows our results.}
	\label{tab:stellar_param}
	\begin{tabular}{@{\,}l@{\,}c@{\,\,\,\,}c@{\,\,\,\,}c@{\,}} 
		\hline
		Parameter & Value & Source & This work\\
		\hline
		$P_{\rm rot}$ (days) & $2.63641\pm 0.00006$ & Hipparcos & $2.63654 \pm 0.00008$ \\
		$L (L_{\odot})$ & $69.35 \pm 11.4$& TIC & \\
		$T_{\rm eff}$ (K) & $11851 \pm 184$ & TIC & 11376 $\pm$ 200 \\
		$\log(g)$ & $4.35\pm 0.15$& TIC & $4.20 \pm 0.3$ \\
		$R (R_{\odot})$ &$1.97 \pm 0.1$ & TIC & \\
		$M (M_{\odot})$ & $3.15\pm 0.1$ & TIC & \\
		$v \sin i$ (km\,s$^{-1}$) &  &  & $35.0 \pm 1.5$ \\
		\hline
	\end{tabular}
\end{table}

These results agree on a parameter spectral type of B9Vp rather than B9IIIp as suggested by the Hipparcos catalog. Note that it is rather common for spectral types of Ap/Bp stars derived from good classification spectra to select luminosity class III rather than V because the H lines are rather narrow for the lower $T_{\rm eff}$ implicitly assumed (Ap rather than Bp classification) because of the very strong Fe and weak or absent He spectrum. Thus, conventional spectral classification for this star could be of type ApCr (no He lines visible, Fe spectral type later than A0, and strong Cr overabundance are shown below).

\subsection{Chemical composition} \label{subsec:chemical_compo}

We need to establish, at least approximately, the abundances of the principal chemical elements that are present in the spectrum, both to more clearly identify what kind of chemically peculiar star V352\,Peg is, and because we need approximate abundances in order to create a suitable line mask to obtain the most precise magnetic measurements possible. There are two main obstacles that prevent a high degree of accuracy in this effort: first, the star is a fairly strong spectrum variable, so (like many magnetic Ap/Bp stars) the abundances of almost all the detectable elements are rather non-uniform over the surface, and therefore the star is not simply characterised by a single abundance value per element; and secondly, the abundances of the main iron group of elements are so large that many weak lines with uncertain atomic parameters crowd the spectrum throughout the visible wavelength window, and make determination of the abundances of less well represented spectra (such as that of Ti or Pr) rather imprecise.

In fact we need to obtain two rather different abundance tables. For the precise measurement of magnetic field strengths using the LSD method, our analysis effectively fits the line absorption and polarisation profile of each spectral line with a simple model, and then averages these fits to obtain a high S/N measurement of circular polarisation. For this purpose we need an abundance table that leads to a linelist that includes lines that approximately reproduce the observed strength of lines when computed with a code that {\it ignores} the effects of a magnetic field, because the selection process for lines in such line data archives as the Vienna Atomic Line Database (VALD) \citep{1995Piskunov,2017Pakhomov} is carried out ignoring the effects of a magnetic field. For specifying the abundances needed to download a suitable linelist, we compute the abundances of the most important elements in the spectrum assuming zero field strength but a modest microturbulence of 2\,km\,s$^{-1}$. This set of abundances will be labelled as set "LL" for "line list". In addition, we want to obtain a different set of abundances, ones that approximately represent the average physical abundance ratios in the atmosphere of V352 Peg, which we must compute taking into account the effect of the stellar magnetic field on spectral line formation, and that consider probable abundance variations over the stellar surface in some way. We will call this second set "MM" for "magnetic model". 

 Abundance fitting for both line lists has been done using the polarised line synthesis program {\sc zeeman} \citep{Landstreet1988,Wade2001,Land11,Baileyetal11}. The program is built around forward computation of the emitted spectrum from an assumed model atmosphere, also assuming an initial abundance table, and if desired a simple model of the stellar magnetic field. The program takes as input a stellar atmosphere interpolated in a grid of ATLAS non-magnetic atmosphere models computed for $T_{\rm eff}$ and $\log g$ values close to those assumed for the star. In its simplest form, the program solves the four coupled equations of transport for polarised radiation propagating through the atmosphere of the specified atmosphere at typically 60 sample points on the visible hemisphere to compute local emergent intensity spectra. These are co-added with appropriate radial velocity shifts (both because of local rotation velocity shifts and due to global radial velocity motion) to predict the observed line spectrum. 
 
\begin{figure*}
   \centering
   \includegraphics[width=18cm,trim={1.0cm 7.0cm 1.8cm 8.0cm},clip]{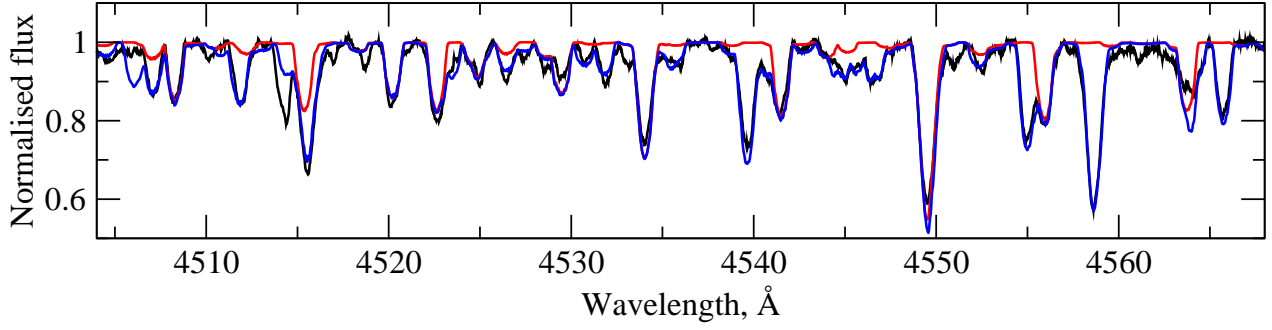}
      \caption{Part of the fit to a spectral window to determine the abundance of Cr. The observed spectrum at phase 0.02 is in black; the fit to the window without Cr is in red, and the best fit to the window after optimising the abundance of Cr, $v_{\rm r}$, and $v \sin i$ is in blue. Spectral features that are sensitive to the abundance of Cr typically have local flux values in the red spectrum near 1.0, but  local flux values roughly following the observed spectrum in the blue spectrum.
              }
         \label{pub-fita-cr}
   \end{figure*}

 This computation uses line lists from VALD, which include both basic atomic data such as $\log gf$ values and damping constants, and also Land{\'e} magnetic splitting factors. The abundances assumed in initial requests for the line lists employed in this project all assumed very large overabundances (relative to solar abundances) of even-Z iron peak elements. The equation of transfer is solved outward in the atmosphere at thousands of wavelengths (spaced by 0.01\,\AA) to predict the emergent spectrum, including at each wavelength the effects of all contributing lines. The program then compares the predicted spectrum with an observed spectrum, and if required, adjusts the abundance table, one element at a time, to find a best fit to a chosen spectral window (see e.g. Fig.\ref{pub-fita-cr}). The stellar radial and rotational velocities are optimised at the same time. We have obtained approximate, but reasonably representative, LL set abundance values by fitting a single spectrum, the spectrum with CFHT odometer number 1331348 from 2011 August 10, which has particularly high $S/N$ data. We assumed a microturbulence of $\xi = 2$ km/s, and made most of the comparisons in $\sim 100$\,\AA\ windows centred at about 4525, 5030, and 6150\,\AA. In this case the normal equation of transfer, rather than the polarised version, is solved. 
 
 In the case that a magnetic field is present, for the physical abundances table MM, the program employs the appropriate atomic data to include magnetic line splitting (the Zeeman effect), computes at each surface grid point the splitting of all relevant spectral lines, and solves the four coupled equations of transfer for polarised radiation to compute both absorption line profiles and line polarisation (Stokes $Q$, $U$ and $V$). In the present application, comparison is made only with the absorption line profiles, again in the same 100\,\AA\ wide windows. At each phase, a single local abundance value is assumed, and possible variations of abundance with optical depth are ignored. When possible, the derived abundance values are checked using more than one spectral window. For each of the elements studied, we have modelled our 100-\AA\ windows at three different rotational phases (0.020, 0.254, and 0.398). Because the deduced abundances from two or three windows at a single phase generally differ by an amount of order 0.2 dex, and similar differences occur from one phase to another, we report a representative compromise abundance for the elements studied. For other elements (O, Mg, Si, Pr, Nd), only one or a few suitable lines are available in the spectrum, but we have again considered three phases. Realistically all of these abundance should be regarded as uncertain by an amount of order 0.2--0.3 dex.

 The resulting abundance values of elements $X$  relative to H, $\epsilon = \log(n_{\rm X}/n_{\rm H})$, are tabulated in Table\,\ref{tab:chemistry}. The second column lists table LL, and the third column table MM. We also list the enhancement or deficiency (in dex) of abundance set MM relative to the solar abundance table as reported by \citet{Asplundetal09}. When a very small number of spectral lines are fit to obtain the abundance, the lines used are noted (by wavelength in \AA\ units) in Table\,\ref{tab:chemistry}. The rare earths Pr and Nd are rather marginally detected in the spectra. Note that no physical abundance is provided for the element O, as the best lines for modelling exhibit very large non-LTE effects that we are unable to correct. 
 
 The dominant peculiarities of this star are the underabundances of He, O, and Mg, and the large overabundances of the iron peak elements Ti, Fe, and especially Cr. However, such abundance anomalies are not unusual for magnetic Ap stars: compare, for example, with the abundance tables found by \citet{Baileyetal11,Baileyetal13} for the slightly hotter magnetic Ap stars HD\,318107 and HD\,147010.

\begin{table}
	\centering
	\caption{Chemical abundances of V352\,Peg. Overabundances and underabundances are computed relative to solar for the MM list.}
	\label{tab:chemistry}
	\begin{tabular}{lcccl} 
		\hline
		Element & $\log(\frac{n_X}{n_H})_{\rm LL}$ &
		$\log(\frac{n_X}{n_H})_{\rm MM}$ & Over/under & lines \\
		        &   (dex)     & (dex) & abund. (dex)  & (\AA) \\
		\hline
		He & $-2.25$  & $-2.5 $ & $-1.4$ & 5876 \\
		O  & $-4.35$  &  --     &  --    & 7772-5 \\
		Mg & $-5.0$   & $-5.3$  & $-0.8$ & 4481 \\
		Si & $-3.75$  & $-4.0$  & $+0.4$ & 5041, \\
		   &          &         &        & 5055-6 \\
		Ti & $-5.35$  & $-5.9 $ & $+1.1$  & many \\
		Cr & $-3.20$. & $-3.8 $ & $+2.5$  & many \\
		Fe & $-2.80$  & $-3.4 $ & $+1.1$  & many \\
		Pr & $-7.30$  & $-6.9 $ & $+4.4$  & 6160-1 \\
		   &          &         &         & 6195 \\
		Nd & $-7.00$  & $-6.5 $ & $+4.0$  &  6145 \\
		\hline
	\end{tabular}
\end{table}

\section{Spectropolarimetric analysis}
\label{sec4}
\subsection{Stokes I and V profiles} \label{subsec:Stokes_IV}

A Least-Square Deconvolution (LSD) is applied to the data, based on the method described in \cite{Donati1997}.
The LSD algorithm requires, as an input, a line mask listing the lines present in the spectra. A template mask was first extracted from VALD3 for $T_{\rm eff}=11500$\,K and $\log g=4.0$ and using the chemical abundances determined above. We used only lines with a depth larger than 1\% of the continuum. We cleaned the mask of hydrogen lines and lines blended with either hydrogen or telluric lines. The final mask contains 5010 lines. We then adjusted the depth of each line in the mask to the observed line depth.
Due to strong line variation between the different spectra, we chose to use a mask with tailored depths per spectrum to obtain more accurate values of the magnetic field. This choice will be discussed in Sect.~\ref{sec:Discussion}.

The circularly polarized spectrum can be written as a convolution of the mean Zeeman signature and the line pattern (see equations presented in \citet{Donati1997} and \citet{Kochukhov2010} for more details).
Polarization information is simultaneously extracted from all the spectral lines present in the mask. 

\begin{figure}
	\includegraphics[width=\columnwidth]{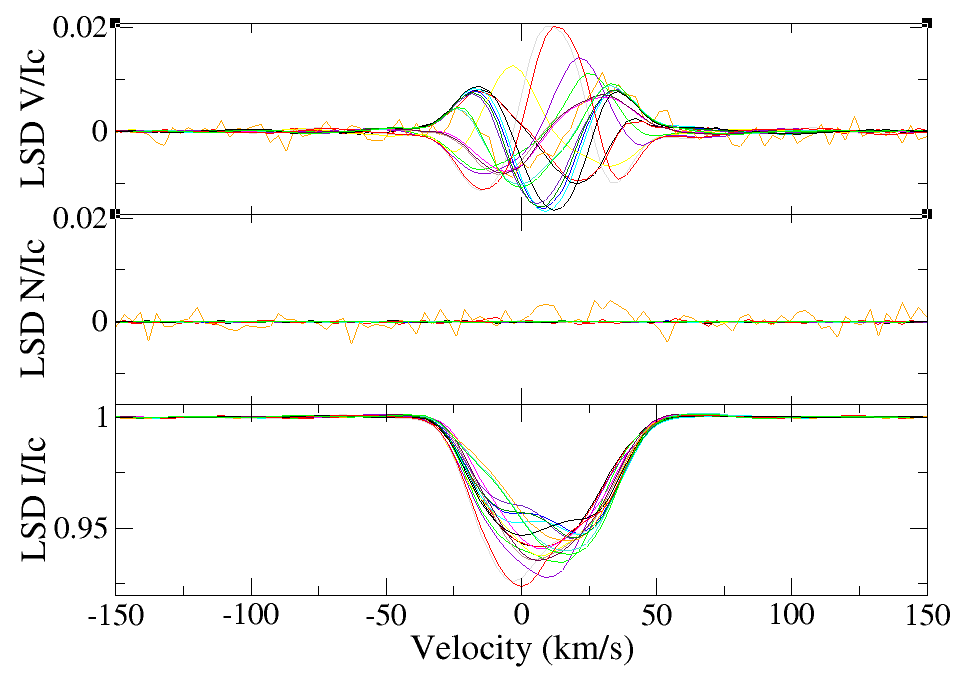}
    \caption{LSD Stokes V (top), Null polarization (middle), and Stokes I (bottom) profiles obtained after the Least-Square Deconvolution of the 18 ESPaDOnS  spectra.}
    \label{fig:LSD}
\end{figure}

The Stokes V and I profiles and Null polarization of V352\,Peg are plotted in Fig.~\ref{fig:LSD}.
Zeeman signatures are visible in all the Stokes V spectra, with an amplitude up to 2\%, while N profiles appear featureless. The Stokes V measurements show variable profiles, typical of a rotational modulation of the magnetic signature. The typical inverse error on Stokes V profiles is 10800.
The Stokes I profiles also show variability, typical of the presence of chemical spots at the surface of the star.

After the deconvolution, the False Alarm Probability (FAP) is computed from the Stokes V and N profiles. This statistical quantity is a detection probability based on the $\chi^2$ statistics.
It is determined both inside and outside the stellar lines and characterizes the presence of a Zeeman signature in the Stokes V profile, enabling to distinguish between a real signature and noise. A detection is classified as definite if the FAP is smaller than $10^{-5}$, corresponding to a detection probability of 99.999\%. The signature is marginally detected if the FAP falls in the interval [$10^{-5}$,$10^{-3}$] and is not detected for a FAP greater than $10^{-3}$.

Inside the stellar lines, the presence of a magnetic field is definitely detected for the 18 observations.
There is also a definite detection for 16 observations outside the stellar lines, due to undulations in the continuum, greater than the noise level. This could be due to small imperfections or errors during the LSD process. An other probable explanation is the residual effects of nearby lines.
The two observations with no detection outside the line profile are the ones with the lowest S/N (observations \#2 and \#11), thus with the higher noise level hiding the undulations.

There is no detection in the N profiles (see Fig.~\ref{fig:LSD}).

\subsection{Longitudinal magnetic field} \label{subsec:Bl}

The longitudinal magnetic field $B_l$ can be computed from the Stokes V and I profiles \citep{Donati1997}, with the following formula:
\begin{equation} \label{eq:Bl}
    B_l = -2.14 \times 10^{11} \frac{\int vV(v)dv}{\lambda g c \int[I_c-I(v)dv]}
\end{equation}
where $\lambda$ and g are the scaling wavelength and Land\'e factor of the stellar lines used to obtain the LSD profile. We used the mean wavelength $\lambda=5406$~\AA\ and the mean Land\'e factor $g=1.19$ to scale the LSD profiles.

The value of $B_l$ varies with the rotation of the star. This is due to a misalignment between the rotation and the magnetic axes, postulated by the Oblique Rotator Model \citep{Stibbs1950}.
For a dipole field, when the stellar magnetic axis is seen ``pole-on'' (pointing in the direction of the observer), the longitudinal magnetic field is at an extremum. On the other hand, when we observe the stellar magnetic equator, $B_l$ is 0 G.

\begin{figure}
	\includegraphics[width=\columnwidth]{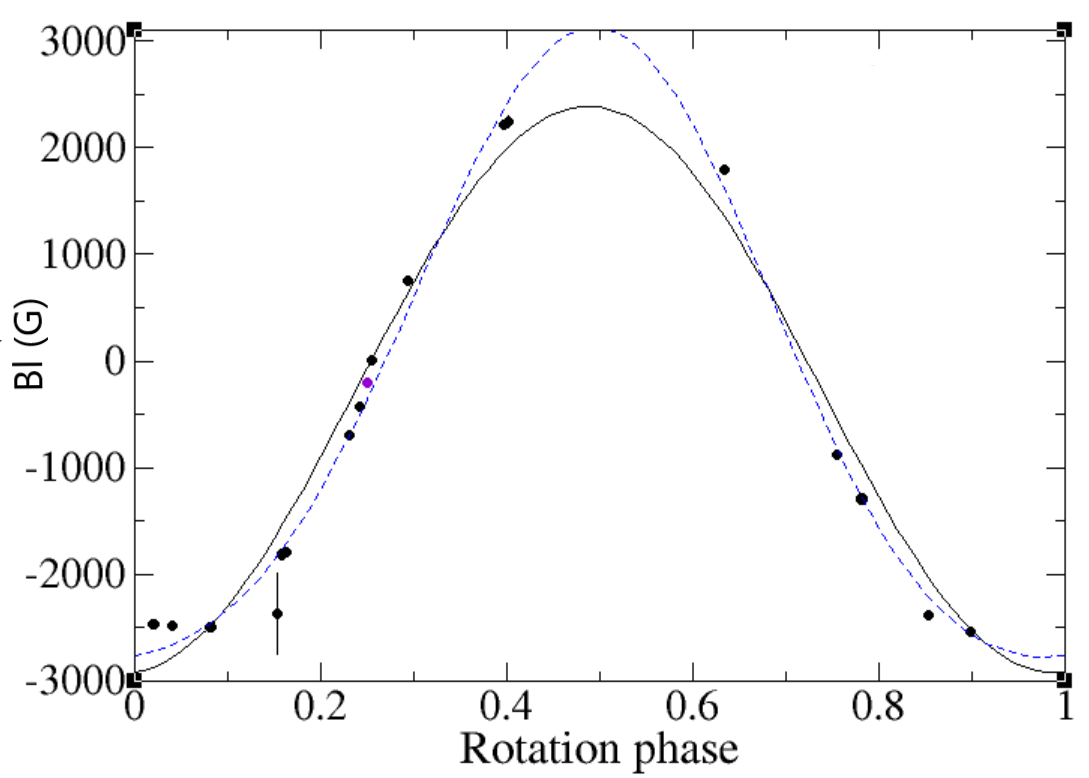}
    \caption{Longitudinal magnetic field of V352\,Peg, folded in phase with HJD$_0 = 2458348.4725$ and P=2.63654 days. The solid black and dotted blue lines are a single-wave and a double-wave fits to the data as defined in Eqs.~\eqref{eq:B_dip} and \eqref{eq:B_quadru}.}
    \label{fig:Bl_plot}
\end{figure}

\begin{table}
	\centering
	\caption{Parameters of the single-wave and double-wave fits to the longitudinal magnetic field values.}
	\label{tab:fit_Bl}
	\begin{tabular}{lc} 
		\hline
		Single-wave & Double-wave \\
		\hline
		$a=-274.23 \pm 5.75$ & $a = -179.02 \pm 0.30$ \\
		$b = 2645.49 \pm 8.26$ & $b = 2943.47 \pm 0.48$\\
		$c = -0.2385 \pm 0.0004$& $c = -0.2395 \pm 2\times 10^{-5}$ \\
		 - & $d = 350.07 \pm 0.38$ \\
		- & $e = 0.2468 \pm 9 \times 10^{-5}$ \\
		Reduced $\chi^2=256$ & Reduced $\chi^2=90$ \\
		\hline
	\end{tabular}
\end{table}

The values of the longitudinal magnetic field for V352\,Peg are presented in Fig.~\ref{fig:Bl_plot}. The data being of very high quality, only one error bar, corresponding to the noisiest spectrum (\#2) is visible.
The black and blue dotted lines in Fig.~\ref{fig:Bl_plot} represent fits to the data described respectively by Eqs.~\eqref{eq:B_dip} and \eqref{eq:B_quadru}:
\begin{align}
    B_l(x) = & a+b\times \sin(x+c) \label{eq:B_dip}\\
    B_l(x) = & a+b\times \sin(x+c) + d\times
    \sin(2x+e)
    \label{eq:B_quadru}
\end{align}
Parameters of the fit are given in Table \ref{tab:fit_Bl}. The errors have been computed from the error on $B_l$.

The variation of the longitudinal magnetic field is better fit with a sine wave described by Eq.  ~\eqref{eq:B_quadru} rather than by Eq. ~\eqref{eq:B_dip}.
The minimum and maximum values of the dipolar fit to the longitudinal magnetic field values are: $B_{\rm l,min} = -2920\pm 15$ G and $B_{\rm l,max} = 2370\pm 15$ G.

\subsection{Rotation period determination from magnetic field measurements} \label{subsec:period}

The variation of the longitudinal magnetic field with time can be used to refine the rotation period.

\begin{figure*}
	\includegraphics[width=1.\linewidth]{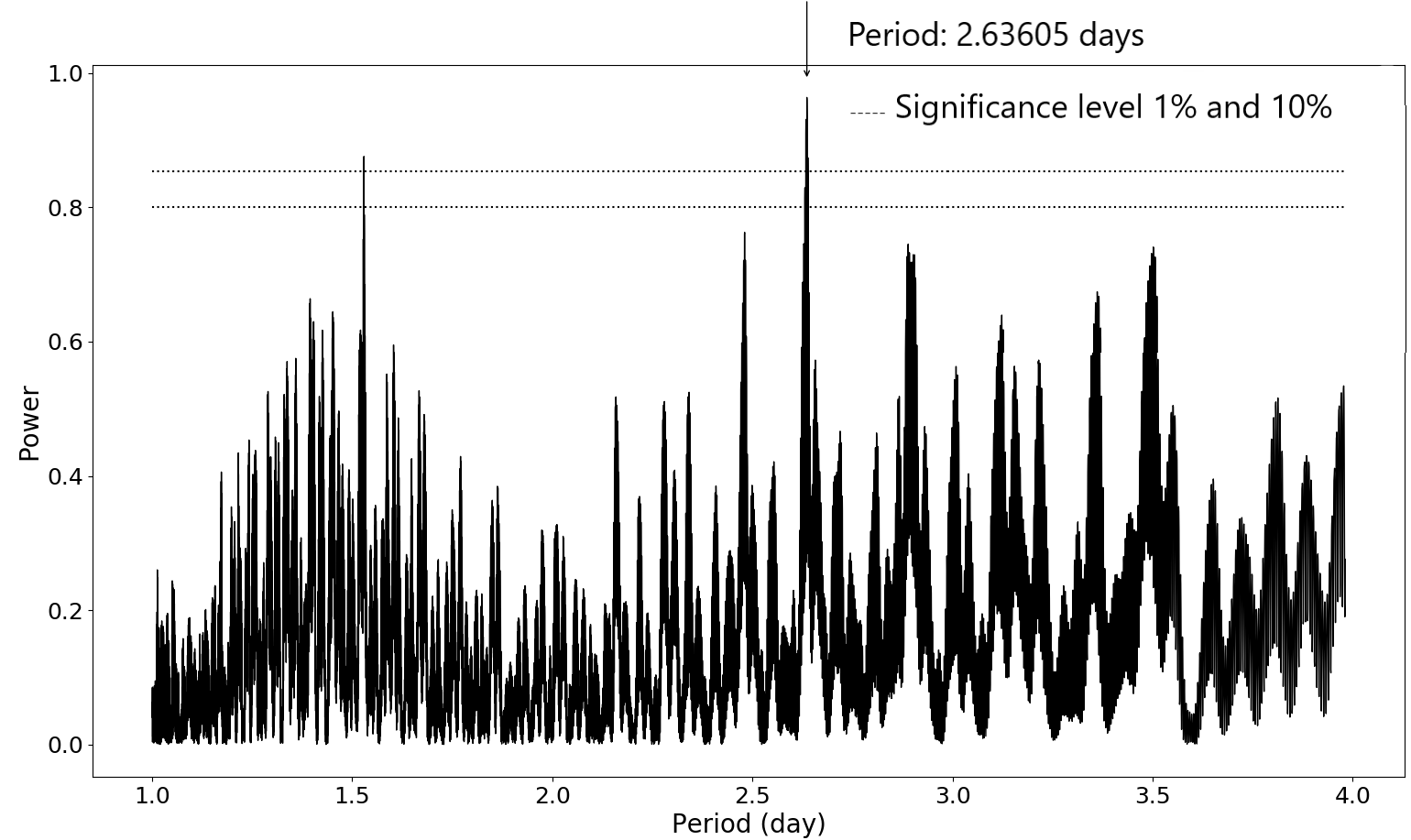}
    \caption{Generalised Lomb-Scargle periodogram of V352\,Peg. The two horizontal dotted lines represent the 1\% and 10\% significance levels.}
    \label{fig:Lomb_periodogram}
\end{figure*}

The observations available are unevenly spaced in time. Instead of the Fourier transform, a more suitable way to find the rotation period is a Least-Squares spectral analysis. This method estimates the frequency spectrum based on a least squares fit of sinusoids to the data set.
Our data set contains only a small number of data points. Thus, a generalised Lomb-Scargle periodogram \citep[GLS,][]{Zechmeister2009} is used to find the peak frequency of the signal. 
Figure~\ref{fig:Lomb_periodogram} has been computed based on a Python implementation of the GLS developed by Jake VanderPlas and reported by \cite{ML}. 

The two dotted lines represent the  significance levels at 1\% and 10\%. Two peaks exceed this threshold: 1.52 and 2.63605 days.
The peak at 2.63605 days is detected with a power of 0.97 and is coherent with the rotation period of 2.64 days presented in the Hipparcos catalog \citep{Hipparcos1997}.
We have no physical explanation for the peak at 1.52 days, but Fig.~\ref{fig:Lomb_periodogram} seems to show a group of incoherent peaks around 1.4-1.5 days so that the 1.52 d value should probably not be considered as a single exact peak.
The longitudinal magnetic field for the rotation period found using the GLS is plotted in Fig.~\ref{fig:Bl_plot_2636}. The fit seems to be in good agreement with the data points. However, by plotting the Stokes I and V profiles (see Sect.~\ref{subsec:Mapping}), we detected a small error on the rotation period: while the observations from August 19, 2011 and December 21, 2018 have a small difference in phase visible in Fig.~\ref{fig:StokesIV_26360}, they have different magnetic signatures. Thus, we used the following reasoning to refine the rotation period.

Between the first observation on the 19th of August 2011 and the mid day of all other observations, the star rotated $N=1012$ times. Thus, a small error on the rotation period will have a strong impact on the phase of the 2011 MiMeS datapoint. Such an error can be estimated based on the misfit between the phase of this point and the expected phase from the dipolar fit.
The expected period phase of V352\,Peg, $\phi_{exp}$, when the 2011 observation was recorded can be calculated from the dipole magnetic field model given by Eq.~\eqref{eq:B_dip}. It is estimated to be $\phi_{exp}=0.08968$. The MiMeS point phase found with the rotation period of 2.63605 days is $\phi_{obs} = 0.90131$. It represents a difference in phase of $\Delta \phi = 1 - \phi_{obs} + \phi_{exp}$, thus an error on the rotation period of $e = \frac{P \times \Delta \phi}{N}=4.9\times10^{-4}$ where $P=2.63605$ days.
The final period we determined is $P_{\rm rot}=2.63654 \pm 0.00008$ days. In the following, all the observations of V352\,Peg are thus phased according to this period and the following reference date HJD$_0 = 2458348.4725$.

\subsection{Dipolar configuration derived from the magnetic field measurements} \label{subsec:analytical_comp}

We model the magnetic field of V352\,Peg as a dipole and use the Oblique Rotator Model, first introduced in \cite{Stibbs1950}. Four parameters are required to describe this model: the inclination angle $i$ between the rotation axis of the star and the line-of-sight, the obliquity angle $\beta$ between the rotation and the magnetic axes, the rotation period $P_{\rm rot}$, and the surface polar field strength of the magnetic dipole $B_{\rm dip}$.

The inclination angle can be constrained from the rotation period and the stellar radius $R$:
\begin{equation} \label{eq:sini}
    \sin{i}=\frac{v\sin{i} \, P_{\rm rot}}{2\pi R}
\end{equation}
From this equation, the inclination angle is estimated to be $i = 66.4\pm 5.4^\circ$. 

The ratio between the minimum and maximum longitudinal magnetic field strength is directly linked to the angles $i$ and $\beta$  \citep{Preston1967}:
\begin{equation} \label{eq:r_ratio}
r \equiv  \frac{B_{\rm l,min}}{B_{\rm l,max}}=\frac{\cos{(i+\beta)}}{\cos{(i-\beta)}}
\end{equation}
Using the values specified in Sect.~\ref{subsec:Bl}, the ratio is estimated to be $r=-1.23\pm 0.01$.
With this ratio and the inclination angle determined above, the obliquity angle is: $\beta = 103.3\pm 7.2^\circ$.

Finally,  $B_{\rm dip}$ can be estimated from the relation \citep{Schwarzschild1950}:
\begin{equation} \label{eq:Blmax}
    B_{\rm l,max} = \frac{15 + u}{15 - 5u}\frac{B_{\rm dip}}{4}\cos({i-\beta}),
\end{equation}
where u, the limb-darkening coefficient, is taken equal to 0.4445 according to \cite{2019Claret}.
We thus obtain a polar magnetic field strength, assuming the field is dipolar, $B_{\rm dip} = 9635\pm1150$ G.

The limb-darkening coefficient has been selected from Table 6 of \citet{2019Claret}, using $T_{\rm eff}=11500$ K, $\log g=4$, and a microturbulent velocity of 4 km/s.
The mean wavelength of the blue passband of Gaia mission being 516.47 nm, close to the mean wavelength of the observations (541.08 nm), we used u1 coefficient from Claret's Table 6 computed using the ATLAS stellar atmosphere model and the least square method, at the Gaia blue wavelength.

\subsection{Mapping the surface magnetic field and relative line strengths}
\label{subsec:Mapping}

\subsubsection{Zeeman-Doppler Imaging model description}

Methods to map the chemical spots of Ap/Bp stars were developped in the 70's, based on line intensity variations in the spectra of magnetic chemically peculiar stars. \citep{1976Khokhlova,1982Goncharskii}. Based on the same principles, \citet{vogt_penrod_1983} introduced the Doppler Imaging method, enabling to obtain a resolved image of a star with temperature spots from spectral line profile distortions.
The Zeeman-Doppler Imaging (ZDI) method was introduced in the 80's \citep{1983Piskunov,1984Piskunov,Semel1989}.
It relies on a similar idea as the Doppler Imaging (DI) technique but applied to circularly polarized data.
DI uses the Doppler effect, applied to rotationally broadened spectral lines, and achieves spatial resolution by mapping the wavelength position along a spectral line to the spatial position across the stellar disk.
An inhomogeneous surface brightness affects the shape of the spectral lines. A darker area produces a bump in the intensity line while a brighter spot generates a depression.
The position of this deformation is modulated by the rotation of the star: as the star rotates, the darker or brighter region propagates across the visible hemisphere of the star and the associated bump or depression propagates across the line profile.
Similarly, a difference in the magnetic field across the stellar surface affects the shape of the spectral lines in total intensity and polarization. Through a modelling of Stokes I and V profiles, Zeeman-Doppler Imaging aims at reconstructing the vectorial magnetic field at the stellar surface.
\newline
ZDI can be performed using individual lines \citep{2014Silvester} as well as LSD profiles \citep{2014Kochukhov,2017Kochukhov}.
An alternative method to reconstruct the surface magnetic field of hot stars is the use of parametrised magnetic field and chemical spots models \citep{2012Bailey,Baileyetal13}.
Linear polarization data (Stokes Q and U) can be added, enabling to recover more complex field typologies and detailed magnetic field maps \citep{2014Silvester}.

Reconstructing a brightness or magnetic map based on a set of line profiles is an ill-posed inverse problem with multiple acceptable solutions. To solve this, one needs an additional regularization constraint, we use the maximum entropy method \citep{Skilling1984}.

The magnetic field topology can be decomposed in spherical harmonics, according to the following set of equations \citep{Donati2006}:
\begin{equation} \label{eq:Br}
    B_r(\theta,\phi)= {\sum_{l=0}^{l_{max}}\sum_{m=0}^{l_{max}}}\alpha_{l,m}Y_{l,m}(\theta,\phi)
\end{equation}
\begin{equation} \label{eq:Btheta}
    B_{\theta}(\theta,\phi)=  - {\sum_{l=0}^{l_{max}}\sum_{m=0}^{l_{max}}}\beta_{l,m}Z_{l,m}(\theta,\phi)+\gamma_{l,m}X_{l,m}(\theta,\phi)
\end{equation}
\begin{equation} \label{eq:Bphi}
    B_{\phi}(\theta,\phi)= - {\sum_{l=0}^{l_{max}}\sum_{m=0}^{l_{max}}}\beta_{l,m}X_{l,m}(\theta,\phi)-\gamma_{l,m}Z_{l,m}(\theta,\phi)
\end{equation}
where $\theta$ and $\phi$ are the colatitude and longitude at the stellar surface.
$X_{l,m},Y_{l,m}$ and $Z_{l,m}$ are the spherical harmonic modes and their derivatives, depending on the Legendre polynomial.
The parameters $\alpha_{l,m}$ and $\beta_{l,m}$ characterize the poloidal component of the magnetic field and $\gamma_{l,m}$ the toroidal part.
The summation is carried out over positive spherical harmonic order $m$, and over degree $l$ varying from 0 to $l_{max}$. The choice of $l_{max}$ is discussed in Appendix \ref{sec:lmax}.

The stellar surface is modelled by a grid of equal area surface elements.
From the above spherical harmonic coefficients, the magnetic field at the element $j$ is evaluated to obtain the vector $[B_r(\theta_j,\phi_j), B_{\theta}(\theta_j,\phi_j),B_{\phi}(\theta_j,\phi_j)]$. Based on that vector and the local line strength $b_j$, the local line profile $\textbf{F}_{j}=(I_{j},V_{j})$, gathering the two local Stokes parameters, is calculated.
Finally, using the set of $\textbf{F}_{j}$ and integrating over the surface of the star, one can get the final line profile $\textbf{F}=(I,V)$. Thus, by knowing $\alpha_{l,m}$, $\beta_{l,m}$, $\gamma_{l,m}$, and $b_j$, and given a line profile model, one can calculate the local line profiles $F_j$ and then obtain the desired disk integrated Stokes spectra: this is the forward problem. The inverse problem consists of reconstructing a magnetic configuration and brightness spots, based on an observed time series of Stokes I and V spectra. This is solved by iteratively fitting a model Stokes I and V profiles to the observations to minimize the $\chi^2$, subject to the additional regularization constraint of maximizing entropy in the brightness and magnetic field distributions. 

The code used to perform the mapping is based on \cite{Folsom2018}. This is a Python implementation of the code presented by \cite{Donati2006} and uses the maximum entropy method of \cite{Skilling1984}, which both minimizes the $\chi^2$ and maximizes the entropy to obtain a unique solution.  
We have made one major modification to the code of \cite{Folsom2018}.  
In Ap stars the Stokes I variability is predominantly driven by chemical spots, which produce variations in line strength, while the DI of \cite{Folsom2018} was based on mapping brightness variations due to temperature spots.  These effects have some similarities, but brightness variations do not change the equivalent width of a line while chemical spots do.  
Thus we have modified the code to map relative line strength, as a proxy for chemical abundance, instead of relative brightness.  The implementation of this modification is relatively straightforward.  The local Stokes I line profile becomes:
\begin{equation}
    I_j(\lambda) = a_{j} d_{j} (1 - b_{j} u(\lambda) )
\end{equation}
for a local line strength $b_j$, using a Voigt function $u$ for wavelength $\lambda$, with a limb darkening $d_{j}$, and projected surface area $a_{j}$, for surface element $j$.
Then the disk integrated, continuum normalized profile is:
\begin{equation}
I(\lambda)/I_c = \frac{\sum_{j} I_j(\lambda)}{\sum_j a_{j} d_{j}}  \label{Ilambda}
\end{equation}
where $j$ is summed over visible surface elements.
The Stokes V local line profile is calculated using the weak field approximation, and also incorporates the distribution of local line strengths. That becomes: 
\begin{equation}
    V_j(\lambda) = -g_{\rm eff} \lambda_0^2 c_0 B_{l,j} \frac{\mathrm{d} I_j(\lambda)}{\mathrm{d} \lambda}
    = g_{\rm eff} \lambda_0^2 c_0 a_{j} d_{j} b_{j} B_{l,j} \frac{\mathrm{d} u(\lambda)}{\mathrm{d} \lambda}
\end{equation}
where $B_{l,j}$ is the line of sight component of the magnetic field for surface element $j$, $g_{\rm eff}$ and $\lambda_0$ are the effective Land\'e factor and wavelength of the line (or the normalizing values for an LSD profile), and the constant $c_0 = \frac{e}{4\pi m c^2}$ using the electron charge $e$, mass $m$, and speed of light $c$.
Then the disk integrated continuum normalized Stokes $V$ profile is:
\begin{equation}
V(\lambda)/I_c = \frac{\sum_{j} V_j(\lambda)}{\sum_j a_{j} d_{j}} . \label{Vlambda}
\end{equation}
This provides a model that incorporates the variation in line strength (equivalent width) produced by chemical abundances spots for both Stokes I and V.  We caution that there is not a linear mapping between the relative line strength mapped here and specific surface abundance values, since line equivalent width depends non-linearly on abundance. However, areas of relatively higher or lower line strength do correspond to areas of higher or lower chemical abundance.

The local line profile model is a Voigt profile, a convolution of a Gaussian and a Lorentzian profile. 
Theoretically, this provides the combination of thermal and turbulent broadening with natural and collisional broadening. Practically, a simple Gaussian is often insufficient to accurately reproduce the Stokes I line shape at high resolution, and since such errors could lead to artifacts in a map, a Voigt profile is preferable. 
The Gaussian width was set to 5.9 km/s to account for the thermal and turbulent broadening and the Lorentzian width to 0.6 km/s. The local Lorentzian and Gaussian widths both have a relatively large uncertainty.  They are also potentially influenced by strong systematic effects such as Zeeman broadening and imperfections in the LSD process, so the values should be treated with caution if interpreting them physically. However, realistic uncertainties on these parameters have a small impact on the derived maps, mostly just changing the contrast of small scale features. We find that taking the Gaussian width in the interval [4 km/s,7 km/s] and the Lorentzian width in [0 km/s,2 km/s] yields to approximately the same results, with a slightly better $\chi^2$ reached for the adopted set of widths. The instrumental broadening is accounted for after the disk integration by convolving the disc-integrated line profile with a Gaussian instrumental profile with R=65000, as described in \citet{Folsom2018}. 
The scaling wavelength and Landé factor have been taken equal to the mean values from the LSD calculation, i.e. $\lambda=540.6$ nm and $g = 1.19$. The parameter controlling the entropy slope, the effect of which is explained in Appendix \ref{sec:E_slope}, is taken equal to 200. The highest spherical harmonic degree $l_{\rm max}$ used to compute the magnetic field is 10. These two parameters have been chosen accordingly to tests presented in Appendices  \ref{sec:E_slope} and \ref{sec:lmax}.

Two statistical tests are used to assess the quality of the ZDI solution.
First, the $\chi^2$ statistics attests of the goodness of the fit, i.e. whether the simulated Stokes I and V profiles match the observed one. An ideal fit has a reduced $\chi^2 = 1$. In this study, we used the approximation that the reduced $\chi^2$ is equal to the $\chi^2$ divided by the total number of observed points.
The second test is introduced in \cite{Skilling1984}:
\begin{equation} \label{eq:TEST}
    {\rm TEST} = \frac{1}{2}\left | \frac{\nabla S}{\left | \nabla S \right |} -  \frac{\nabla C}{\left | \nabla C \right |} \right  |^2
\end{equation}
TEST measures how anti-parallel the gradients in entropy and $\chi^2$ are.
The entropy used is defined in \citet{Folsom2018} by Eqs.~B11 and B13, introduced respectively in \citet{1998Hobson} for the magnetic spherical harmonic coefficients and \citet{Skilling1984} for the relative line strengths.
A fully converged image should have TEST=0.
Thus, the suitable magnetic and line strength configuration should ideally have $\chi^2=1$ and TEST=0.

\subsubsection{Refinement of the stellar parameters with ZDI} \label{subsec:refinement_param}

\begin{figure}
\centering
\includegraphics[width=1\linewidth]{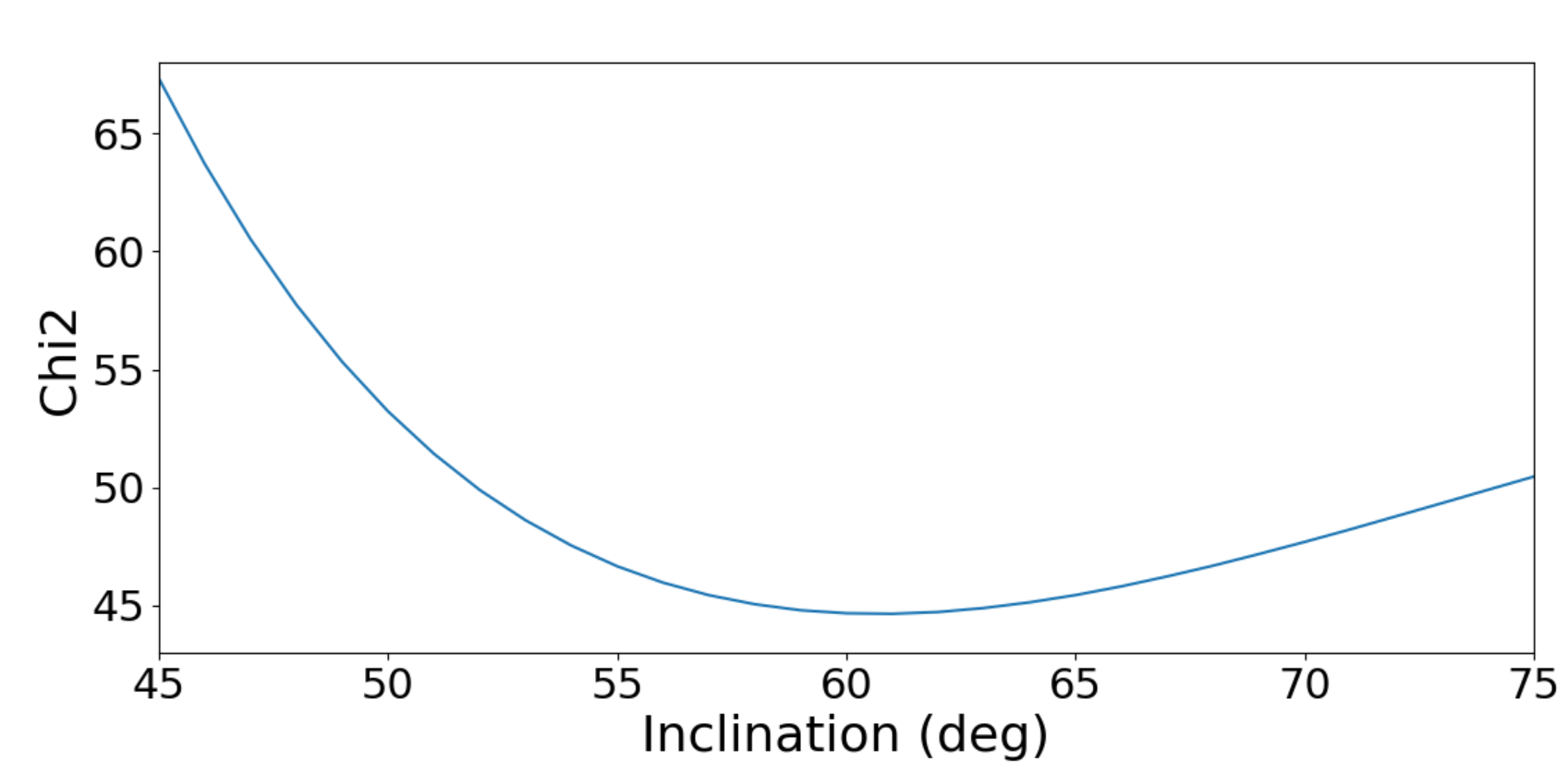}
\caption{Refinement of the value of the inclination angle $i$ with a target entropy $E=-180000$.}
\label{fig:Inc_fixed_E}
\end{figure}

The inclination $i$ and the projected rotation velocity can be refined using ZDI, based on the method presented in \citet{2002Petit}.
A grid of inclination angles and $v\sin i$ values is generated.
A target entropy is set while $\chi^2$ remains unconstrained.
For each value of the inclination angle, the ZDI algorithm is run to fit the observed Stokes V profiles. The result of this method is presented in Fig.~\ref{fig:Inc_fixed_E}.
The same method is used to refine the projected rotation velocity but Stokes I profiles are used instead of V in this case, because Stokes V is much less sensitive to $v\sin i$ than Stokes I (and conversely for the inclination).
The optimal parameter is the one minimizing $\chi^2$ for a given target entropy. Another procedure yielding approximately the same result consists in setting a target $\chi^2$ value and a free entropy. The optimal parameter is then the one maximizing the entropy for a fixed $\chi^2$.

\cite{Folsom2018} described a method to compute the error on the parameters based on the significance of variations in the $\chi^2$ statistic.
The variation in the $\chi^2$ about the minimum corresponding to a $1\sigma$ confidence level provides the range of the reported error in $i$ (or $v\sin i$). 
After computation of the error, the inclination is set to $i={62^\circ} ^{+9}_{-7}$, agreeing with the value of  $i={66.4^\circ} \pm 5.4$ derived from the longitudinal field curve and presented in Sect.~\ref{subsec:analytical_comp}, within error bars.
The same procedure is used to refine the projected rotation velocity. A value of $v\sin i=35.0 \pm 1.5$ km/s is adopted. We note that Zeeman splitting is not explicitly included in the Stokes I line model, this could lead to an overestimate of vsini, or of the local line width, and both parameters may be overestimated by a few km/s.

\subsubsection{Magnetic and relative line strength configuration of V352\,Peg}

\begin{figure*}
\centering
\includegraphics[height=11cm]{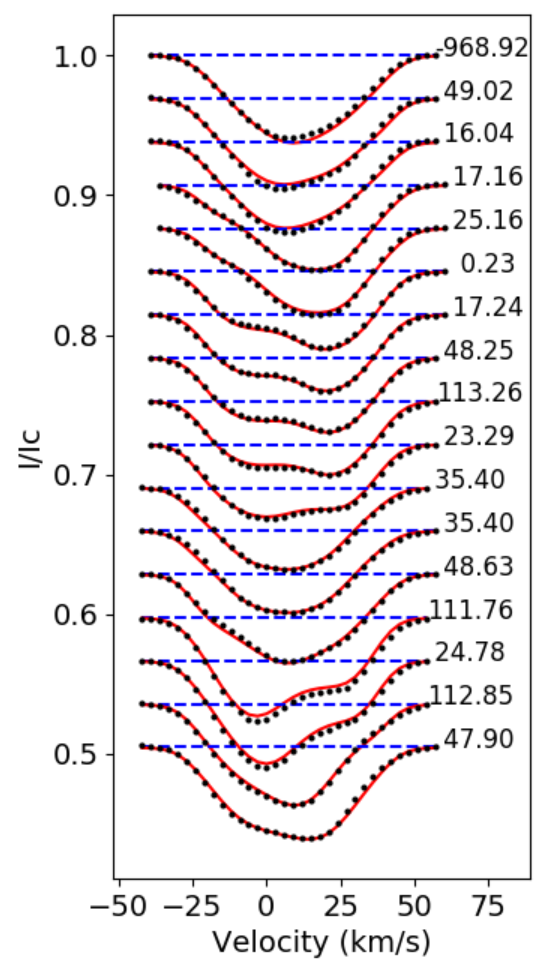}
\includegraphics[height=11cm]{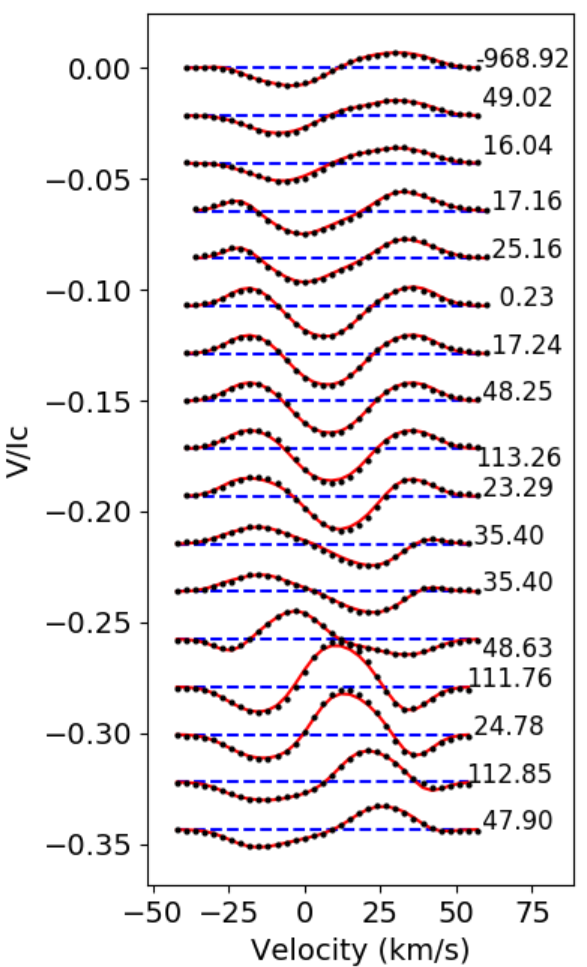}
\caption{Comparison between observed (black dots) and simulated (red lines) Stokes I (left panel) and V (right panel) profiles. The profiles are shifted vertically according to their rotation phase, which is indicated next to each profile.}
\label{fig:StokesIV}
\end{figure*}

The results of the fit to Stokes I and V profiles using the optimized stellar parameters are plotted in Fig.~\ref{fig:StokesIV}.
The goodness of the fit is assessed by the $\chi^2$ value and TEST parameter. 
Stokes I profiles were fitted to a reduced $\chi^2$ of 100 while the final reduced $\chi^2$ from Stokes V fit is 40. 
These values might appear high, the optimal value being 1. This can be explained because the error bars on the Stokes I and V LSD profiles are extremely small.

\begin{figure*}
\centering
\includegraphics[width=1\linewidth]{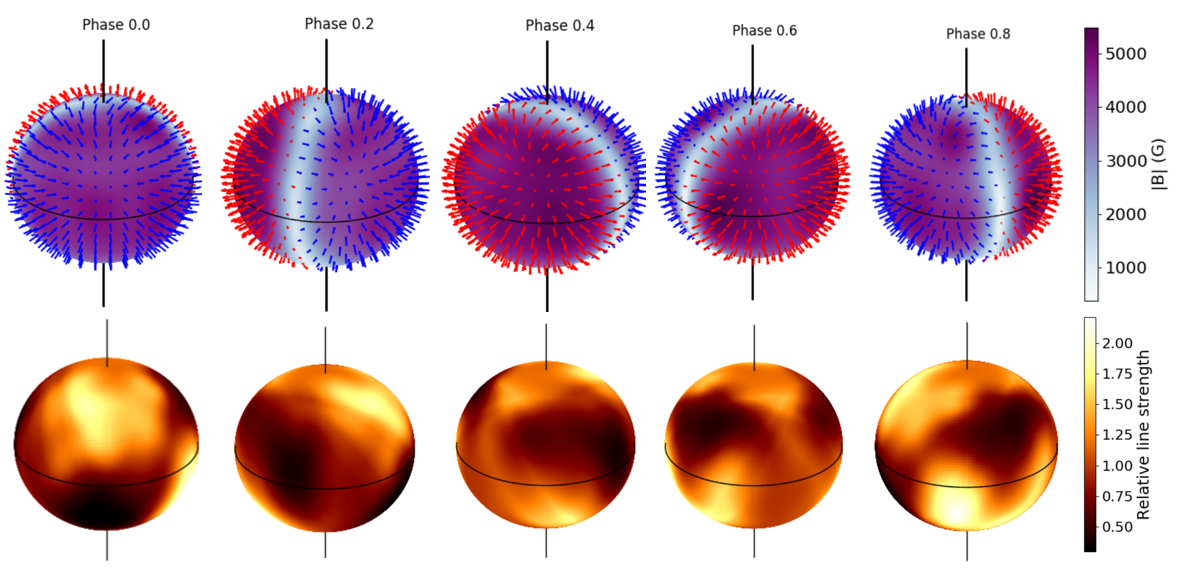}
\caption{Magnetic (top row) and relative line strength (bottom row) maps of V352\,Peg. Each column corresponds to a different rotation phase. The black axis is the rotation axis. The colors represent the magnetic field strength in Gauss (top row) and the relative line strength (bottom row). The arrows at the stellar surface indicate the vectorial magnetic field (red arrows have a positive radial field and blue arrows have a negative radial field).}
\label{fig:Magsmooth_phase}
\end{figure*}

2D surface reconstruction of V352\,Peg for five different rotation phases are plotted in Fig.~\ref{fig:Magsmooth_phase}. The magnetic equator is visible, almost aligned with the rotation axis. The obliquity angle between the magnetic and rotation axes is estimated to be $\beta=91.5 \pm7.0^{\circ}$. Like the inclination angle, the obliquity angle estimated with ZDI agrees with the estimate (see Sect.~\ref{subsec:analytical_comp}) of $\beta = 103.3\pm 7.2^\circ$ derived from the longitudinal field curve, within the error bars. The polar magnetic field strength is $B_{\rm dip} = 6320^{+170}_{-280}$ G, differing more from the analytical estimate. This difference is discussed in Sect.~\ref{sec:Discussion}. Both $B_{\rm dip}$ and $\beta$ have been estimated from the radial poloidal $l=1$ component of the magnetic field using ZDI. The error on these two parameters has been computed by propagating the error on $i$ in Eqs.~\eqref{eq:r_ratio} and \eqref{eq:Blmax}. 

\begin{figure*}
\centering
\includegraphics[width=1\linewidth]{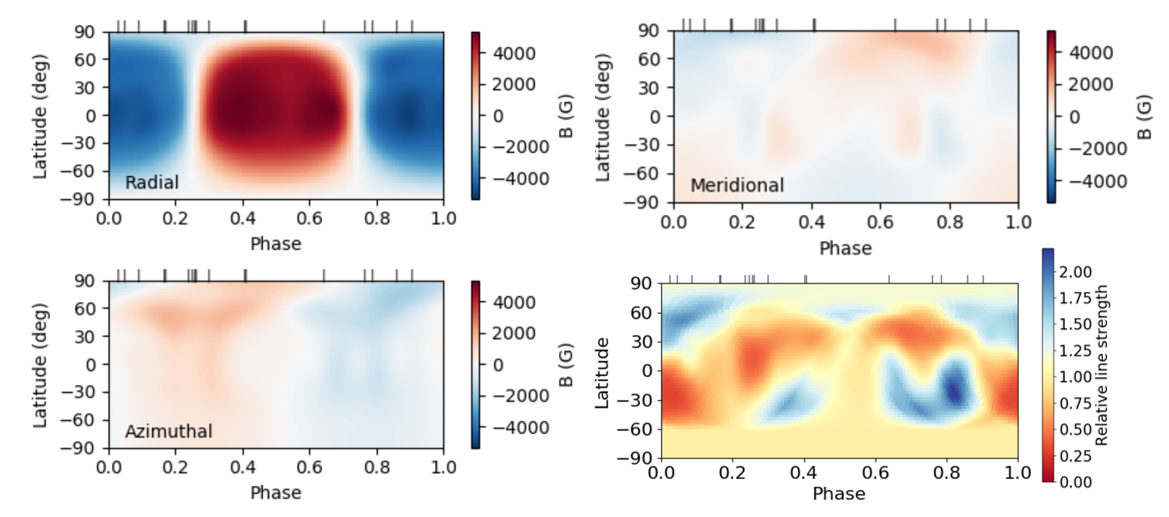}
\caption{ZDI map for V352\,Peg. The vertical lines above each map indicate the phases of individual spectra. The figure on the bottom right represents the relative line strength.}
\label{fig:BrightMagPolar}
\end{figure*}

From Figs.~\ref{fig:Magsmooth_phase} and \ref{fig:BrightMagPolar}, the magnetic field is clearly dipolar.
However, there is a clear asymmetry in the strength of the magnetic field between the positive and negative regions. The magnetic field strength reaches 5200 G in the positive hemisphere while its maximum in the magnetic negative  hemisphere is 4900 G (in absolute value). Moreover, the magnetic field strength is not homogeneous within a magnetic hemisphere, as attested by spots of stronger magnetic field visible in the radial magnetic field map in Fig.~\ref{fig:BrightMagPolar}.

The relative line strength map is shown in the bottom row of Fig.~\ref{fig:Magsmooth_phase}. Two large spots of high relative line strength are visible at $\phi=0$ in the Northern hemisphere and at $\phi=0.8$ in the Southern hemisphere. In addition, the magnetic equator, observable at the rotation phases 0.2, 0.4, 0.6, and 0.8, seems to coincide with regions of lower relative line strength (thus lower absorption), except in the polar regions.
The relative line strength configuration does not exactly follow the dipole configuration of the magnetic field. This is also visible in the polar projection maps in Fig.~\ref{fig:BrightMagPolar}.

\begin{figure*}
\centering
\includegraphics[width=1\linewidth]{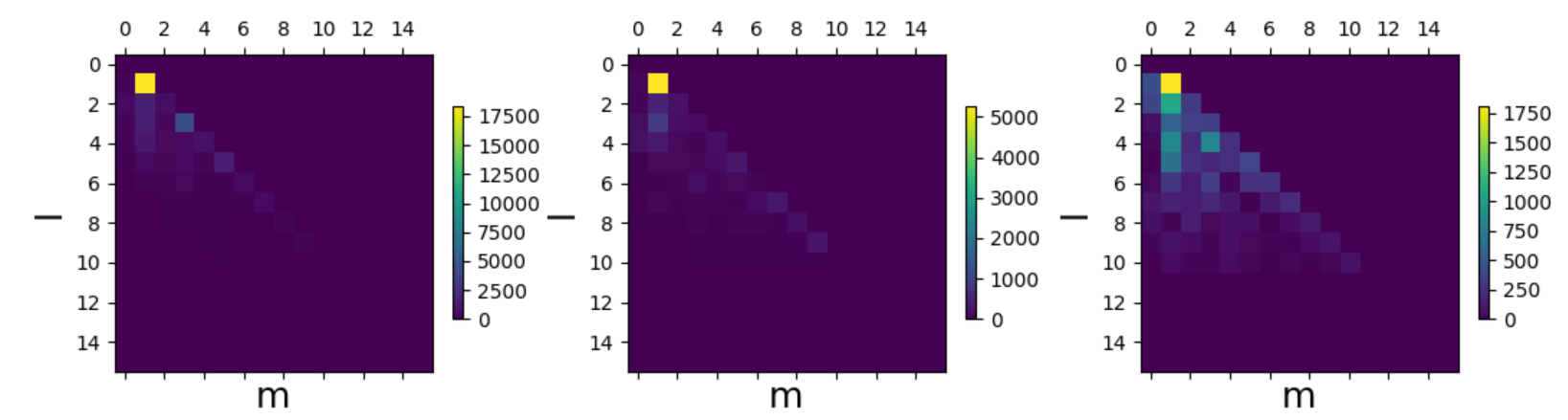}
\caption{Modulus of the spherical harmonic complex coefficients as a function of mode degree l and order m. $\alpha_{l,m}$, $\beta_{l,m}$, and $\gamma_{l,m}$ correspond respectively to the
radial field component, tangential component of potential field term, and tangential component of toroidal field term.}
\label{fig:sph_harm}
\end{figure*}

The magnetic field energies in different components of the geometry are described in Table~\ref{tab:mag_confi} and taken to be proportional to $B^2$.
An other way of seeing the configuration is by plotting the modulus of the spherical harmonic coefficients.
This is done in Fig.~\ref{fig:sph_harm}. The axisymmetric modes correspond to $m=0$ spherical harmonics.
\begin{table}
	\centering
	\caption{Magnetic field energies obtained through ZDI.}
	\label{tab:mag_confi}
	\begin{tabular}{lc} 
		\hline
		Magnetic component & Percentage \\
		\hline
		 Poloidal & $98.5\%$ (\% tot) \\
		 Toroidal & $1.5\%$ (\% tot) \\
		 Axisymmetric & $0.4\%$ (\% tot) \\
		 Dipole & $91.1\%$ (\% pol) \\
		Quadrupole & $1.0\%$ (\% pol) \\
		Octopole & $5.9$ (\% pol) \\
		Poloidal axisymmetric & $0.31\%$ (\% pol) \\
		Toroidal $l=1$ & $31.9\%$ (\% tor) \\
		Toroidal $l=2$ & $18.1\%$ (\% tor)\\
		Toroidal $l=3$ & $7.7\%$ (\% tor)\\
		Toroidal axisymmetric & $7.0\%$  (\% tor)\\
		Dipole axisymmetric & $0.1\%$  (\% dip)\\
		\hline
	\end{tabular}
\end{table}

The magnetic field is mainly poloidal, with a very weak toroidal component contributing to  only 1.5\% of the total magnetic energy.
The poloidal field is dominantly dipolar (91.1\%) and non-axisymmetric, with weak contributions from the quadrupolar and octupolar components. 
For the coefficients $\alpha_{l,m}$ and $\beta_{l,m}$, the dominant mode is $l=1$ and $m=1$, corresponding to a non-axisymmetric dipole.
Contrary to the poloidal field, the toroidal part is characterized by a stronger contribution from higher order multipoles (essentially quadrupole and octupole with $m=1$ and $l=2,3$). However, since the toroidal component is very weak, it is not clear how reliable the geometry is. 
Modes with $l>6$ contribute faintly to the image reconstruction, so do axisymmetric modes ($m=0$), especially for the poloidal part of the magnetic field.

\subsubsection{Surface distribution of Titanium, Chromium and Iron} \label{subsubsec:chemical_maps}
Surface distribution maps can be obtained for Titanium, Chromium and Iron (elements with a high number of spectral lines, see Table~\ref{tab:chemistry}).
\newline
The surface distribution of an element can be inferred from the variation of local line strength (or local equivalent width) \citep{1989Hatzes,1991Hatzes}. A region enhanced in a chemical element will create deeper absorption features in the spectra, increasing the local equivalent width (proportional to the local line strength). Thus, by modelling the local line strength variation, we can obtain information about the surface distribution of chemical elements.
\newline
As mentioned in \cite{1989Hatzes}, going from a local equivalent width to actual abundances is not trivial. In this section we aim at approximating the surface distribution of an element in term of stronger or weaker local line strength, that should correlate with enhanced or depleted abundances.
\newline

Spectral lines from Cr, Fe, and Ti are used to reconstruct the surface distribution of these elements.
To this aim LSD profiles are produced from a subset of lines extracted from the original LSD line mask (see Sect.~\ref{subsec:Stokes_IV}). 
We used around 2000 lines of Cr, 3000 lines of Fe, and 190 lines of Ti.

\begin{figure*}
\centering
\includegraphics[width=\linewidth]{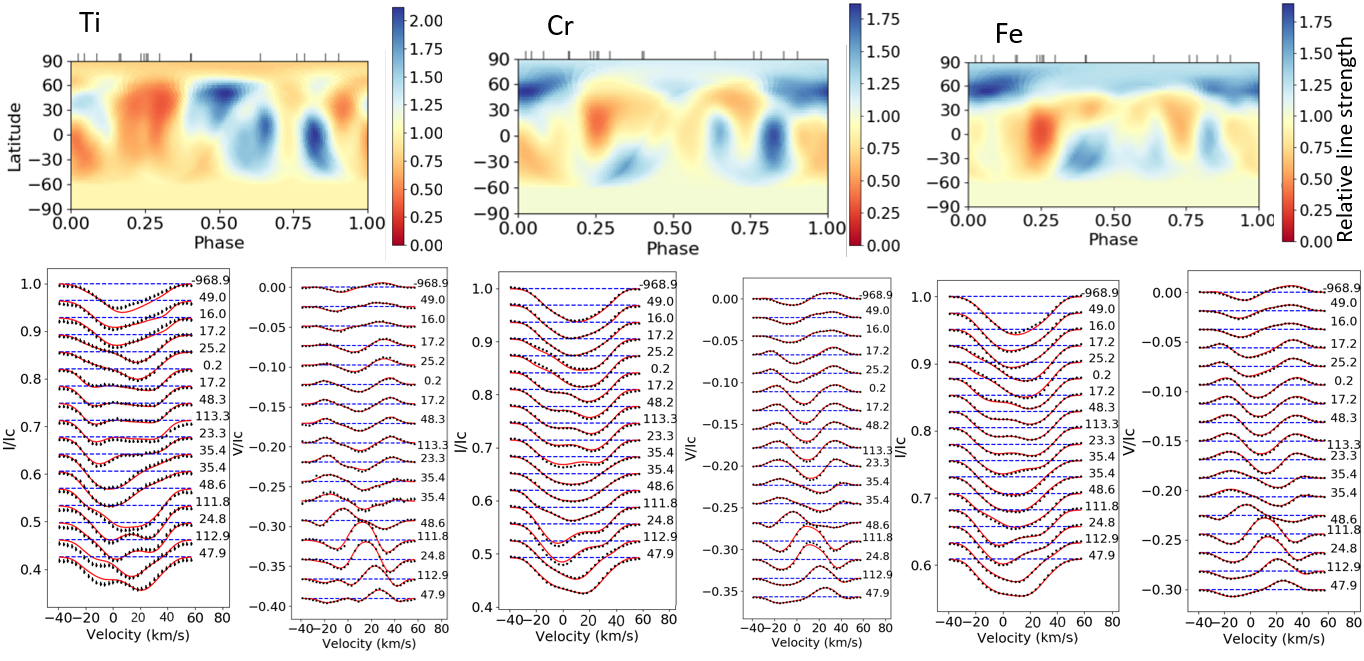}
\caption{Relative line strength map of Ti, Cr, and Fe. The corresponding fit to Stokes I and V profiles is shown below each map.}
\label{fig:chemical_spots}
\end{figure*}

Results are presented in Fig.~\ref{fig:chemical_spots}. The surface distribution of Cr, Fe, and Ti is inhomogeneous. As expected from the clear variations observed in the line profiles, chemical spots are present at the surface of V352\,Peg. The Stokes I and V profiles are very well fitted for Iron and Chromium. For Titanium, the ZDI code struggles to fit the profiles at phases 112.85, 47.90 and -968.92. These phases should coincide with the Southern magnetic hemisphere entering in the line-of-sight, around phase 0.9 in the map from Fig.~\ref{fig:chemical_spots}.
\newline
By comparing Figs.~\ref{fig:BrightMagPolar} and \ref{fig:chemical_spots}, one can see that Ti seems to gather around the magnetic equator, visible at the phase 0.75 in Fig.~\ref{fig:BrightMagPolar}. The distribution of Cr and Fe is very similar and more complicated to link to the magnetic topology. There is an overabundant region located in the Northern polar region. The overall distribution of these two chemical elements could be linked to the meridional magnetic field map in Fig.~\ref{fig:BrightMagPolar} but there is no clear trend.

\section{Discussion}
\label{sec:Discussion}

In this study we have used the LSD method, which assumes that the magnetic field is weak and that the lines considered in the line  deconvolution have a similar profile. 
The weak field approximation (WFA) might be a questionable assumption as the polar magnetic field strength we derived in the ZDI model reaches $B_{\rm dip} = 9.6$ kG. \cite{2004Landi} suggested that the weak field regime breaks down around 1 kG for iron-peak spectral lines. In addition, \citet{Kochukhov2010} showed that the LSD single line approximation becomes inaccurate with increasing magnetic field strength.
Using the WFA also has an impact on the modelling with ZDI.
Thus, we designed a test to check for the validity of the WFA.

We calculated line profiles for a dipole magnetic field with strength of 500 G, 5 kG, and 10 kG using {\sc zeeman} \citep{Landstreet1988} based on a model \ion{Fe}{II} line, and assuming a homogeneous iron abundance. Details about the line data are presented in Appendix \ref{sec:WFA_test}. We then treated the line profiles as LSD profiles, added synthetic noise, and used them to reconstruct the magnetic field using the ZDI code by \citet{Folsom2018}. We also studied the impact of two input parameters on the reconstruction of the magnetic topology with the ZDI code: $l_{\rm max}$, setting the maximum degree of the spherical harmonic expansion describing the magnetic field, and $E_{\rm slope}$, controlling the change in slope in the entropy curve. More information about these tests can be found in Appendices  \ref{sec:E_slope} and \ref{sec:lmax} along with tables gathering the results. In this discussion, we only address the WFA.

The reconstructed magnetic topology for the three values of dipolar field strengths is presented in Table \ref{tab:WFA_validity}. The magnetic topology is almost perfectly recovered for a dipole field of 500 G and 5 kG, and very well recovered for the 10 kG field, only departing from the true topology by less than 3\%. In the three cases, the obliquity angle $\beta$ is well estimated within $\pm 2^\circ$. On the other hand, the dipolar magnetic field strength is slightly underestimated for a magnetic field of 500 G (by 7\%), with larger underestimates for larger fields (23\% for 5 kG and 28\% for 10 kG). We can conclude that the WFA enables to recover the magnetic topology very well even for a field of 10 kG. The main effect appears to be an underestimation of about 30\% of the polar magnetic field strength at 10 kG. Therefore, the value of $B_{\rm dip} = 6320^{+170}_{-280}$ G is underestimated by about 30\%, leading to a more realistic polar magnetic field strength of $\sim$ 9 kG. This value is indeed in better agreement with the analytical estimate 
$9635\pm 1150$ G found in Sect.~\ref{subsec:analytical_comp}. Regarding the magnetic field topology of V352~Peg, the poloidal component is dominant and we find that the tangential component of the poloidal field is significantly weaker than the radial component (generally $\alpha_{l,m} > \beta_{l,m}$). The tangential components of a magnetic field are often less well constrained than the radial component in ZDI maps based only on Stokes $V$ observations. However, in the synthetic test in Appendix B, a stronger tangential poloidal component is reconstructed than in the observations, which suggests this is not purely a systematic error.  Thus we tentatively conclude that the tangential poloidal component is weaker than the radial component, although Stokes $Q$ and $U$ observations could help confirm this.

A particularity of this spectropolarimetric analysis has been to tailor one line mask per spectrum on the LSD analysis. Due to the high variation of the spectral lines between the different spectra, the mask should be adapted to each spectrum for a correct computation of the magnetic field. Thus, based on one cleaned template mask, 18 masks have been generated to fit the spectral lines of the 18 spectra. 
The longitudinal magnetic field computation and ZDI analysis have been carried out both using one single and 18 tailored masks in order to analyze the impact and consequences of this choice on the results.
The shapes of the LSD profiles are similar. The main difference is the width of the profiles. The single mask we used has been tailored to spectrum \#3 (2297797), taken close to the magnetic equator, when the magnetic field is weaker. Consequently, the magnetic signature extracted from other spectra obtained "pole-on" is underestimated.
As the longitudinal magnetic field is computed from Stokes signatures according to Eq.~\eqref{eq:Bl}, the change in line depth impacts the value of $B_l$. However, this change is very weak (points are shifted only by a few Gauss). 
Finally, the ZDI code run on LSD files computed from a single mask shows similar results as the one presented in Sect.~\ref{subsec:Mapping}. The parameters computed following Sect.~\ref{subsec:refinement_param} are not impacted by this mask change. 

\section{Conclusion}

In this paper, we presented the first spectropolarimetric analysis of the magnetic chemically peculiar hot star V352\,Peg. With an effective temperature of $T_{\rm eff} = 11400$ K and gravity $\log g = 4.22$ found with Geneva photometry in good agreement with the TESS input catalog, this is an early main sequence star. Chemical abundances and the projected rotation velocity have been derived using the {\sc zeeman} code from \citet{Landstreet1988}. Chromium, iron, and titanium are found in large excess while oxygen, helium and magnesium are depleted. We have thus reclassified the star as a B9Vp star.

Stokes I and V profiles were obtained through a least-square deconvolution of the spectra. Zeeman signatures are clearly detected in all the spectra, with a clear modulation of the signature with the rotation phase due to a misalignment between the rotation and magnetic axis of the star, predicted by the Oblique Rotator Model. 
Stokes I profiles also show a variability revealing the presence of chemical spots on  the stellar surface.
The modulation is also visible in the longitudinal magnetic field strength $B_l$, varying with the rotation phase of the star. This variation has been used to refine the rotation period with a Lomb-Scargle periodogram. Then, using the archival observation from 2011, the rotation period could be refined even more, leading to the very good accuracy of $P_{\rm rot} = 2.63654 \pm 0.00008$ days.

Zeeman-Doppler Imaging implemented by \cite{Folsom2018} has been used to obtain magnetic and brightness maps of the stellar surface as well as to refine the geometrical parameters.
We find $i={62^\circ}^{+9}_{-7}$, $v\sin i=35.0 \pm 1.5$ km/s, and $\beta=91.5^{\circ} \pm 7.0$ in good agreement with the analytical computation. Using ZDI, the polar magnetic field strength has been estimated to be $B_{\rm dip} = 6320^{+170}_{-280}$ G, differing from the analytical value by $\sim$35\%. 
When testing for the validity of the WFA, we found that the  estimate of $B_{\rm dip}$ by ZDI is underestimated by up to 30\% for strong fields. We thus conclude that the polar magnetic field strength of V352\,Peg is close to $\sim$9 kG. 
This underestimation is the main effect of the WFA on the reconstructed magnetic topology. In synthetic tests of the WFA, the relative magnetic energies of the poloidal and toroidal component as well as the dipole percentage of the poloidal field are well recovered within 3\% for a 10 kG field and the magnetic field maps presented in Appendix~\ref{sec:WFA_test} are very close to the true topology. We also tested the impact of the magnetic field on the reconstructed relative line strength map of the star (see Appendix~\ref{sec:WFA_test}). We find that the overall relative line strength is well recovered. For a magnetic field strength of 5 kG and greater, regions of enhanced and reduced relative line strength are visible. Such regions trigger relative line strength variation of about $20\%$ with respect to the expected value. This effect, although reasonable, should be taken into account when analysing the stellar relative line strength map.

The magnetic field topology of V352\,Peg is mainly dipolar, which is consistent with usual configurations for magnetic Ap and Bp stars. The toroidal component is very weak, and the magnetic field highly non-axisymmetric with respect to the rotation axis. 
Surface distribution of Chromium, Iron, and Titanium have been derived. Cr and Fe seem to have similar distributions with an enhanced region located in the Northern pole and spots at latitude $-20^\circ$. The surface distribution of Ti is differing, with an enhanced and large spot visible at phase 0.5.

Additional spectropolarimetric observations in linear polarization (Stokes Q and U) could enable to refine the surface distribution of these chemical elements and obtain maps of higher precision.

\section*{Acknowledgements}
{We thank the referee for his careful reading of the manuscript and his suggestion to modify the code of \cite{Folsom2018} to map relative line strengths. Based on observations obtained at the Canada-France-Hawaii Telescope (CFHT) which is operated by the National Research Council(NRC) of Canada, the Institut National des Sciences de l’Univers of the Centre National de la Recherche Scientifique (CNRS) of France, and the University of Hawaii. This research has made use of the SIMBAD database operated at CDS, Strasbourg (France), of NASA’s Astrophysics Data System (ADS), and of the VALD database, operated at Uppsala University, the Institute of Astronomy RAS in Moscow, and the University of Vienna.}

\section*{Data Availability}
{The spectropolarimetric data are publicly available in the Polarbase database at http://polarbase.irap.omp.eu.}



\bibliographystyle{mnras}
\bibliography{v352peg_ref} 


\appendix

\section{Refinement of the rotation period} \label{sec:annex_P}

\begin{figure}
	\includegraphics[width=\columnwidth]{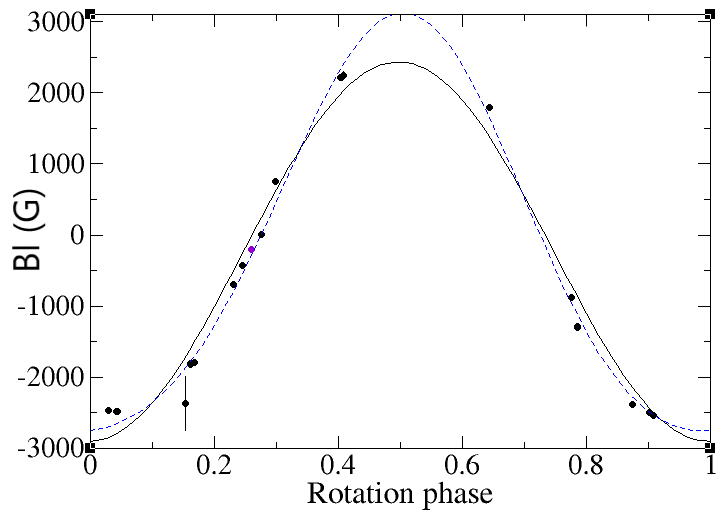}
    \caption{Longitudinal magnetic field of V352\,Peg, folded in phase with HJD$_0 = 2458348.4725$ and P=2.63605 days. The black and blue lines are a single-wave and a double-wave fits to the data.}
    \label{fig:Bl_plot_2636}
\end{figure}

\begin{figure}
\centering
\includegraphics[width=0.46\columnwidth]{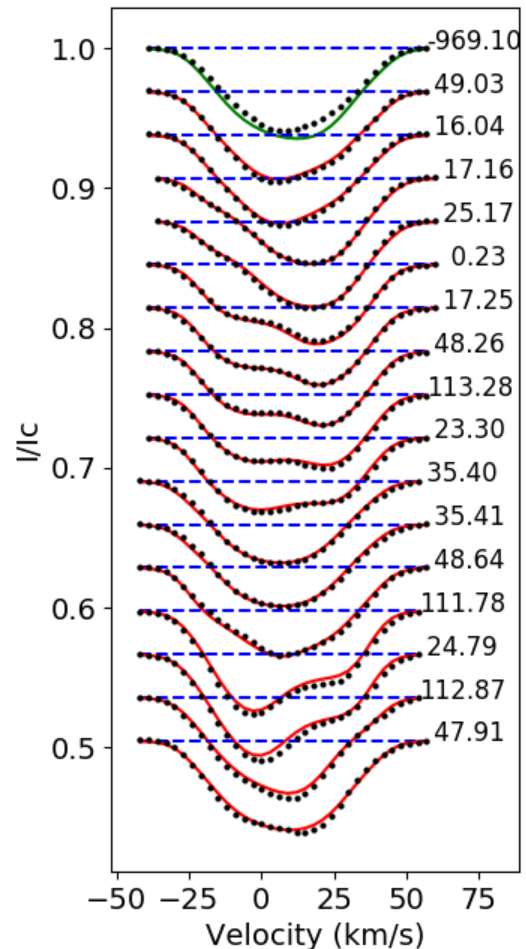}
\includegraphics[width=0.49\columnwidth]{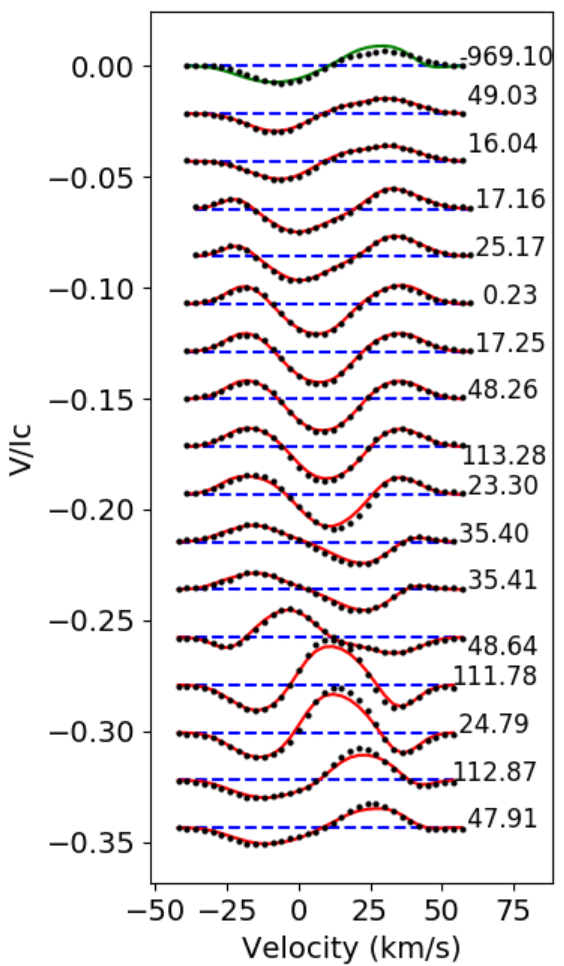}
\caption{Comparison between observed (black dots) and simulated (red lines) Stokes I (left panel) and V (right panel) profiles computed with HJD$_0=2458348.4725$ and P=2.63605 days. Stokes profiles from the MiMeS observation in 2011 are plotted in green  and show that the period used here is not precise enough.}
\label{fig:StokesIV_26360}
\end{figure}

Using the MiMeS datapoint from 2011, we could refine the rotation period estimated using a Lomb-Scargle (LS) periodogram.
Fig.~\ref{fig:Bl_plot_2636} represents the longitudinal magnetic field of V352\,Peg as a function of the rotation phase for the period of 2.63605 d found using a LS periodogram.
In Fig.~\ref{fig:Bl_plot_2636}, the data point from 2011 is located at a phase of 0.901, very close to the observation from the 21st of December 2018 (with a phase of 0.908) and to the single-wave and double-wave fits. 
However, the Stokes profiles show different signatures, which is not expected for two points relatively close in phase as can be seen from the ZDI modelling (Fig.~\ref{fig:StokesIV_26360}). Using this, we refined the rotation period as mentioned in Sect.~\ref{subsec:period}.

\FloatBarrier

\section{Test of the validity of the weak field approximation} \label{sec:WFA_test}

\begin{figure*}
\centering
\includegraphics[width=2\columnwidth]{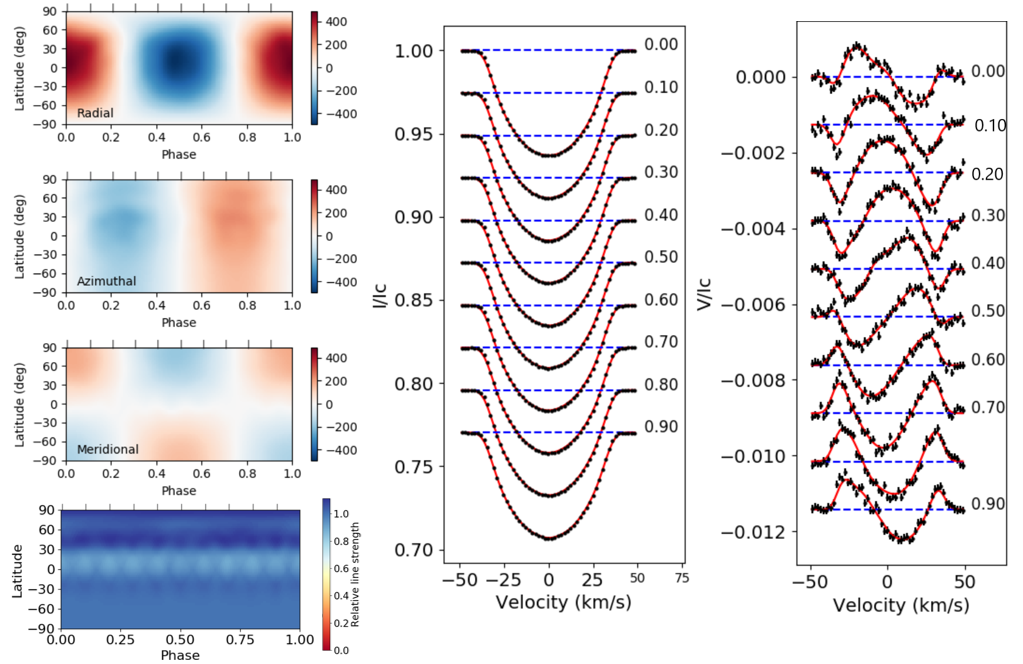}
\caption{Left panel: reconstructed magnetic field maps (three first rows) and relative line strength (bottom row). The middle and right panels show the corresponding fit to Stokes I and V profiles for an input dipole magnetic field strength of 500G.}
\label{fig:all_500}
\end{figure*}

As mentioned in Sect.~\ref{sec:Discussion}, we designed a test to check for the impact of the WFA on ZDI results. We generated synthetic line profiles for dipolar fields with strengths of 500 G, 5 kG, and 10 kG using {\sc zeeman}, and then attempted to fit them using the ZDI code. {\sc zeeman} does not rely on the WFA and uses full Zeeman splitting calculations and polarized radiative transfer. Weaker field values are used to check the accuracy of the magnetic maps for a modest magnetic field, while the stronger field values are used to quantify the impact of departures from the WFA. A model \ion{Fe}{ii} line with a wavelength of 5000 \AA\ was used in {\sc zeeman} along with J quantum numbers for the lower and upper levels of 1 (for a simple triplet pattern) and Land\'e factors for the lower and upper level of 1.2.  The $\log {gf}$ was -1.0,  the lower level excitation potential was 5.3 eV, and the radiative, quadratic Stark, and van der Waals damping parameters were respectively 8.3, -5.7, and -7.6.  These line parameters were chosen because they are  typical of a line in the visible wavelength domain for a 10000 K star. The stellar angles are taken to $i=60^\circ$ and $\beta=90^\circ$. The other stellar parameters are chosen close to the one of V352\,Peg ($T_{\rm eff}=10000$ K, $\log g = 4.5$, $v\sin i = 35$ km/s). Note that the test is very weakly sensitive to the temperature.
We then added synthetic noise to the line profile. As the amplitude of Stokes V changes a lot with the field strength, we scaled the noise level with the magnetic field strength. We used a noise level of 0.0001 for 500 G, 0.001 for 5 kG, and 0.002 for a 10 kG field strength. The noise level is taken relative to the continuum (as $1 \sigma$ of the Gaussian distribution).

\begin{figure*}
\centering
\includegraphics[width=2\columnwidth]{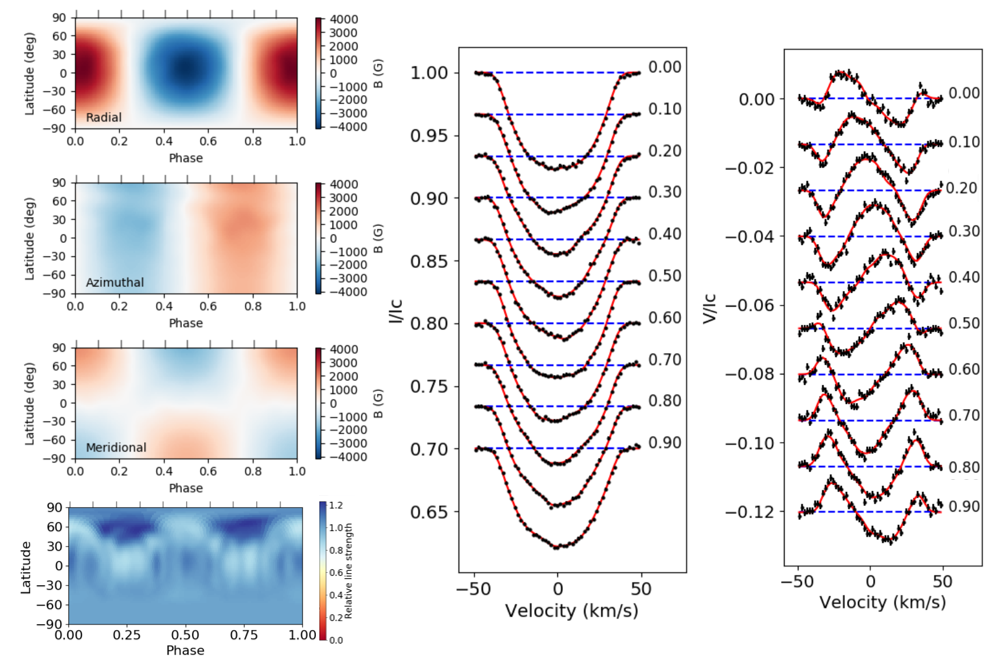}
\caption{Same as Fig.~\ref{fig:all_500} for an input dipole magnetic field strength of 5000G.}
\label{fig:all_5000}
\end{figure*}

\begin{figure*}
\centering
\includegraphics[width=2\columnwidth]{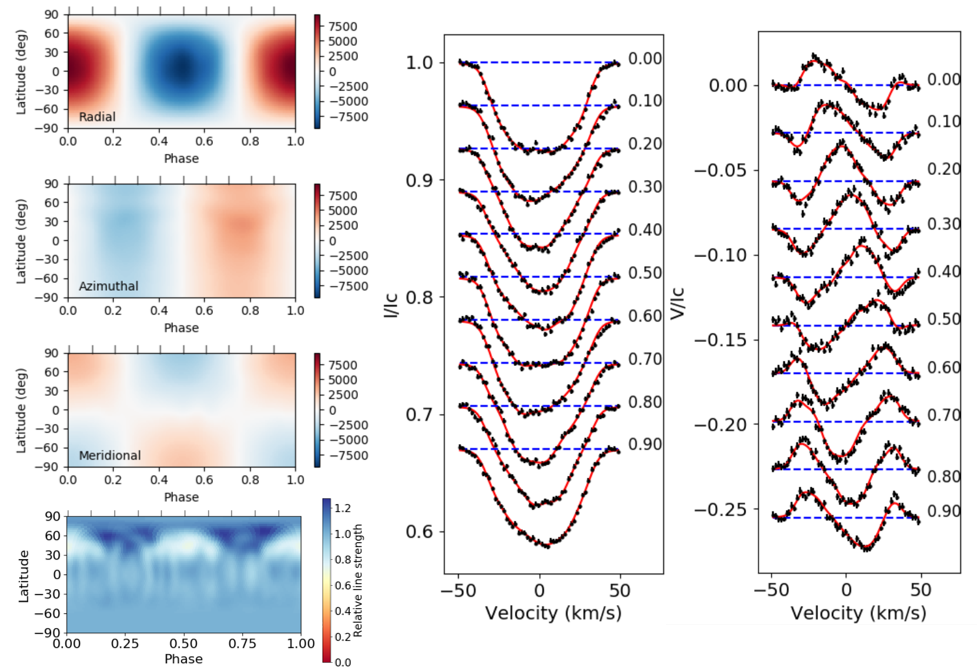}
\caption{Same as Fig.~\ref{fig:all_500} for an input dipole magnetic field strength of 10 kG.}
\label{fig:all_10000}
\end{figure*}

For the ZDI reconstruction we adopted the same $v\sin i$, $i$, rotation phases, line wavelength, and effective Landé factor as in the line model. The local Gaussian and Lorentzian line widths were set by fitting the Stokes I profiles through $\chi^2$ minimization, with the final values of 4.5 and 0.2 km/s, respectively.  We then ran the ZDI code on the synthetic line profiles to obtain a magnetic configuration for the three values of dipolar magnetic field. The relative magnetic energies and dipole magnetic field strength estimated using ZDI are gathered in Table \ref{tab:WFA_validity} while magnetic and relative line strength maps and fits to Stokes I and V profiles are presented in Figs.~\ref{fig:all_500}, \ref{fig:all_5000} and \ref{fig:all_10000}. These results are discussed in Sect.~\ref{sec:Discussion}.

\begin{table} 
	\centering
	\caption{Reconstructed magnetic topology for different input dipolar magnetic field strengths ($B_{\rm dip}$).}
	\label{tab:WFA_validity}
	\begin{tabular}{lccc} 
		\hline
		Input $B_{\rm dip}$ (G)& 500 & 5000 & 10000\\
		Noise level used & 0.0001 & 0.001 & 0.002\\
		\hline
		$B_{\rm dip}$ (G) & 464 & 3847 & 7206 \\
		Error from true value & 7\% & 23\% & 28\% \\
		Poloidal (\% tot)& 99.7 & 99.7 & 98.2 \\
		Dipole (\% pol) & 99.5 & 99.3 & 97.4 \\
		\hline
	\end{tabular}
\end{table}


We find that the magnetic topology is very well reconstructed for all three strengths, with a few percent error at 10 kG, and less than 1\% error at lower field strengths.  The magnetic field strength of the dipolar component is only slightly underestimated at 500 G (by 7\%), while it is more significantly underestimated at 5 kG (23\%) and 10 kG (28\%). Thus we conclude that moderate violations of the WFA can have a moderate impact on the reconstructed field strength, which may be important for some applications, while the magnetic geometry is still robustly reconstructed. 

It is also interesting to study the impact of a stronger field on the reconstructed relative line strength map, supposed to be homogeneous. For a 500 G, the map in Fig.~\ref{fig:all_500} is close to homogeneous, with two positive variations of $+0.1$ in relative line strength at latitude 90$^{\circ}$ and  50$^{\circ}$ and an underestimation of the relative line strength at the equatorial region. As the strength of the magnetic field increases, symmetric patterns appear in the relative line strength map. In particular, two dark spots arise around the phases 0.25 and 0.75 at a latitude of 60$^\circ$. A bright spot is also visible between latitude 0 and 60$^\circ$, moving along the stellar surface as the star rotates. For a dipole magnetic field strength of 5 kG (Fig.~\ref{fig:all_5000}) and even for 10 kG (Fig.~\ref{fig:all_10000}), the effect of increasing magnetic field strength in the reconstructed relative line strength map is still moderate.

\section{Effect of the entropy slope parameter on the ZDI results} \label{sec:E_slope}

Our ZDI methodology relies on a maximum entropy regularization, and the entropy formulation requires one control parameter, which we call $E_{\rm slope}$.
$E_{\rm slope}$ controls the change in slope in the entropy curve and can be defined based on Eq.~\eqref{eq:S} corresponding to Eq.~(8) in \citet{1998Hobson} when $m_f = m_g = E_{\rm slope}$. 
\begin{equation}\label{eq:S}
    S[\textbf{h},\textbf{m}_f,\textbf{m}_g]=\sum_{i}(\psi_i -(m_f)_i -(m_g)_i-h_i\ln[\frac{\psi +h_i}{2(m_f)_i)}])
\end{equation}
where $\psi_i = [h_i^2 + 4(m_f)_i(m_g)_i]^{\frac{1}{2}}$.

Eq.~\eqref{eq:S} generalises Eq.(2.2) from \citet{1989Skilling} for the entropy of a positive distribution \textbf{f} to a positive/negative distribution $\textbf{h}= \textbf{f} - \textbf{g}$, with $\textbf{m}_f$ and $\textbf{m}_g$ being two separate models for $\textbf{f}$ and $\textbf{g}$, acting like prior information. For the spherical harmonic description of the magnetic field, the values $\textbf{h}$ are the spherical harmonic coefficients $\alpha$, $\beta$, and $\gamma$ \citep[e.g.][]{Folsom2018}. The real and imaginary parts of the coefficients are included as separate real terms in the summation. Here, we chose $\textbf{m}_f = \textbf{m}_g$ since we consider positive and negative magnetic fields with equal probability. Thus, $\textbf{m}_f = \textbf{m}_g  =  E_{\rm slope}$ and this parameter gives prior information on the distributions $\textbf{f}$ and $\textbf{g}$. Being directly linked to the entropy by Eq.~\eqref{eq:S}, $E_{\rm slope}$ influences the regularization of the magnetic field. Spherical harmonic coefficients below that value contribute less to the entropy so they are getting less strongly regularized. If the value of $E_{\rm slope}$ is too large it could lead to the regularization being too weak. On the other hand, if $E_{\rm slope}$ is too small, there is a risk of underfitting. 

We used three values of entropy slope, scaled with respect to the dipolar magnetic field strength. For example, with $B_{\rm dip} = 500$ G, we took $E_{\rm slope} = 1$, 10, and 100, while we multiplied these values by a factor 10 for a field with $B_{\rm dip} = 5$ kG. We then used the ZDI code by \citet{Folsom2018} to reconstruct the magnetic field, using the synthetic line profiles described in Appendix \ref{sec:WFA_test}. We are especially interested in the poloidal and dipolar  percentage of the total magnetic energy (supposed to be 100\%) as well as the reconstructed dipolar magnetic field strength. We fit Stokes V profiles to a reduced $\chi^2=1.3$.
The results are gathered in Table~\ref{tab:B_slope_500}.

\begin{table} 
	\centering
	\caption{Impact of the entropy slope on the reconstruction of a 500 G, 5 kG, and 10 kG magnetic field. The reconstructed field strength, error from the true value (in \%), and pourcentage of poloidal and dipolar components are shown.}
	\label{tab:B_slope_500}
	\begin{tabular}{@{\,}l@{\,\,\,}c@{\,}c@{\,}c|c@{\,}c@{\,}c|c@{\,}c@{\,}c@{\,}} 
		\hline
Field strength & \multicolumn{3}{c|}{500 G} & \multicolumn{3}{c|}{5 kG} & \multicolumn{3}{c}{10kG} \\
		\hline
		$E_{\rm slope}$ & 1 & 10 & 100 & 10 & 100 & 1000 & 20 & 200 & 2000\\
		\hline
		$B_{\rm dip}$ (G) & 490 & 463 & 394 & 4080 & 3835 & 3243 & 7300 & 7291 & 6336 \\
		Error & 2\% & 7\% & 21\% & 18\% & 23\% & 35\% & 27\% & 27\% & 37\%\\
		Poloidal (\% tot)& 100 & 99.7 & 96.7 & 100 & 99.7 & 96.5 & 98.7 & 99.0 & 95.4 \\
		Dipole (\% pol) & 100 & 99.5 & 95.5 & 99.9 & 99.3 & 94.7 & 97.5 & 97.9 & 93.2 \\
		\hline
	\end{tabular}
\end{table}

The trend is similar for the three magnetic field strengths. For a $E_{\rm slope}$ value about the same order of magnitude as the dipolar field strength, the reconstructed magnetic topology is less well recovered than for a smaller value of entropy slope. 
This is because a larger $E_{\rm slope}$ provides a weaker constraint on the magnetic field, allowing more of the magnetic energy to be distributed in small values of higher degree spherical harmonic coefficients.
In the case of a 10 kG field, we see that the poloidal and dipolar percentages increase slightly between a value of $E_{\rm slope}= 20$ and 200. This might be due to underfitting when $E_{\rm slope}= 20$.
The value for the entropy slope that appears to enable a good regularization without overfitting small scale structures is about 50 times weaker than the dipolar magnetic field strength.
However, this value might differ from case to case depending on the complexity of the magnetic field. For complex fields with contribution of higher order multipoles, this value could be higher, maybe of the order of $B_{\rm dip}$. V352\,Peg having a simple field configuration, we used $E_{\rm slope}= 200$ in our analysis.

\section{Effect of the maximum spherical harmonic degree on the ZDI results} \label{sec:lmax}

In the ZDI analysis, $l_{\rm max}$ is the maximum spherical harmonic degree used to compute the magnetic field based on Eq.~\eqref{eq:Br}, \eqref{eq:Btheta}, and \eqref{eq:Bphi}. 
This controls the number of free parameters used in the magnetic field inversion in ZDI.
$l=$1, 2, and 3 correspond respectively to the dipolar, quadrupolar, and octupolar components. 
We tested the impact of $l_{\rm max}$ using the synthetic line profiles from Appendix \ref{sec:WFA_test} with three different magnetic field strengths. 
We reconstructed magnetic maps with $l_{\rm max}=$ 5, 10, and 15, and we fit the Stokes V profiles to a target reduced $\chi^2=1.4$. The results are shown in Table~\ref{tab:lmax_500}.
Overall, the value of $l_{\rm max}$ does not impact the reconstructed magnetic topology, especially for dipolar magnetic field strengths of 500 G and 5 kG. Modes with $l>5$ contribute little to the reconstructed image. 
However, when the magnetic field reaches 10 kG, limiting the spherical harmonic degrees to 5 worsen slightly the results (by 1\%) than limiting them to 10 or 15. Taking this into account, using $l_{\rm max}=10$ appears to be a good compromise.

\begin{table} 
	\centering
	\caption{Impact of parameter $l_{\rm max}$ on the reconstruction of a 500 G, 5 kG, and 10 kG magnetic field.}
	\label{tab:lmax_500}
	\begin{tabular}{@{\,}l@{\,\,\,}c@{\,}c@{\,}c|c@{\,}c@{\,}c|c@{\,}c@{\,}c@{\,}} 
		\hline
Field strength & \multicolumn{3}{c|}{500 G} & \multicolumn{3}{c|}{5 kG} & \multicolumn{3}{c}{10kG} \\
		\hline
		$l_{\rm max}$ & 5 & 10 & 15 & 5 & 10 & 15 & 5 & 10 & 15\\
		\hline
		$B_{\rm dip}$ (G) & 463 & 466 & 460 & 3834 & 3826 & 3814 & 7148 & 7311 & 7325\\
		Error & 7\% & 7\% & 8\% & 23\% & 23\% & 24\% & 29\% & 27\% & 27\%\\
		Poloidal (\% tot)& 99.7 & 99.7 & 99.7  & 99.7 & 99.7 & 99.7 & 98.1 & 99.2 & 99.4 \\
		Dipole (\% pol) & 99.5 & 99.5 & 99.5 & 99.3 & 99.3 & 99.3 & 97.0 & 98.1 & 98.3\\
		\hline
	\end{tabular}
\end{table}


\bsp	
\label{lastpage}
\end{document}